\documentclass[aps,prb,twocolumn,showpacs,notitlepage,floatfix,longbibliography,superscriptaddress]{revtex4-1}

\usepackage{amsfonts}
\usepackage{amsmath}
\usepackage[all]{xy}
\input xy
\xyoption{matrix}
\xyoption{2cell}
\xyoption{arrow}
\xyoption{curve}
\xyoption{all}
\usepackage{lipsum}

\usepackage{graphicx}
\usepackage{physics}

\usepackage[normalem]{ulem}
\usepackage{subfigure}
\usepackage[dvipsnames]{xcolor}

\usepackage{appendix}
\usepackage[breaklinks=true,colorlinks=true,linkcolor=blue,citecolor=blue,filecolor=blue,urlcolor=blue,pdfstartview=FitH]{hyperref}
\usepackage{float}

\newcommand{\pksadd}{Max Planck Institute for the Physics of Complex Systems, Dresden D-01187, Germany}
\newcommand{\pcsadd}{Center for Theoretical Physics of Complex Systems, Institute for Basic Science(IBS), Daejeon 34126, Korea}
\newcommand{\ustadd}{Basic Science Program(IBS School), Korea University of Science and Technology(UST), Daejeon 34113, Korea}
\newcommand{\umassadd}{Department of Mathematics and Statistics, University of Massachusetts, Amherst MA 01003-4515, USA}

\newcommand{\mh}{\mathcal{H}}
\newcommand{\hmh}{\hat{\mh}}
\newcommand{\hmhb}{\hmh}

\newcommand{\ha}{\hat{a}}
\newcommand{\hb}{\hat{b}}
\newcommand{\hc}{\hat{c}}
\newcommand{\hp}{\hat{p}}
\newcommand{\hf}{\hat{f}}

\begin{document}

\title{Quantum Caging in Interacting Many-Body  All-Bands-Flat Lattices}

\author{Carlo Danieli}
\affiliation{\pksadd}
\affiliation{\pcsadd}

\author{Alexei Andreanov}
\affiliation{\pcsadd}
\affiliation{\ustadd}

\author{Thudiyangal Mithun}
\affiliation{\pcsadd}
\affiliation{\umassadd}

\author{Sergej Flach}
\affiliation{\pcsadd}
\affiliation{\ustadd}

\date{\today}

\begin{abstract}

We consider translationally invariant tight-binding all-bands-flat networks which lack dispersion. In a recent work [arXiv:2004.11871] we derived the subset of these networks which preserves nonlinear caging, i.e. keeps compact excitations compact in the presence of Kerr-like local nonlinearities. 
Here we replace nonlinear terms by Bose-Hubbard interactions and study quantum caging.
We prove the existence of degenerate energy renormalized compact states for two and three particles, and use an inductive conjecture to generalize to any finite number M of participating particles in one dimension.
Our results explain and generalize
previous observations for two particles on a diamond chain [Vidal et.al. Phys. Rev. Lett.  85, 3906 (2000)]. 
We further prove that quantum caging conditions  guarantee the existence of extensive sets of conserved quantities in any lattice dimension, as first revealed in 
[Tovmasyan et al Phys. Rev. B 98, 134513 (2018)] for a set of specific networks. 
Consequently transport is realized through moving pairs of interacting particles which break the single particle caging.

\end{abstract}

\maketitle


\section{Introduction}

The study of localization phenomena in systems of interacting particles gave rise to some of most remarkable research streams in condensed matter physics during the past decades. Typically these phenomena arise in the absence of translation invariance, as both the first prediction of single particle localization~\cite{anderson1958absence,kramer1993localization} and the finite temperature transition to many-body localized phases of weakly interacting quantum particles~\cite{basko2006metal,aleiner2010finite} have been obtained in tight-binding networks in the presence of uncorrelated spatial disorder. However, both single and many-body particle localization can be achieved in translationally invariant setups ({\it e.g.} see Refs.~\onlinecite{pino2016nonergodic,abanin2019colloquium} for discussion on disorder-free many-body localization). 

One of the notable examples of single particle localization in translationally invariant lattices are \emph{flatband} networks -- lattices where at least one of the Bloch energy bands is independent of the wave vector (hence, dispersionless or flat).~\cite{derzhko2015strongly,leykam2018artificial,leykam2018perspective}
Differently from Anderson localization where the disorder turns all the eigenstates exponentially localized, 
the eigenstates associated to a flat band have strictly compact support spanned over a finite number of unit cells -- and therefore they are called compact localized states (CLS).
Since their first appearance as mathematical testbeds for ferromagnetic ground states ~\cite{mielke1991ferromagnetic,tasaki1992ferromagnetism} the extension of the flatband concept to novel lattice geometries has been of crucial interest. This gave rise to several generator schemes~\cite{ dias2015origami,morales2016simple,maimaiti2017compact,rontgen2018compact,toikka2018necessary,maimaiti2019universal} adopting different principles to design artificial flatband lattices in different spatial dimensions.  

If \emph{all} Bloch bands are flat, then \emph{all} single particle eigenstates are spatially compact. In this all bands flat (ABF) case, single particle transport is fully suppressed and non-interacting particles remain caged within a finite volume of the system. This caging phenomenon due to the collapse of the Bloch spectrum was originally dubbed {\it Aharonov-Bohm} (AB) caging since it emerged from a finetuning of a magnetic field into a time-reversal symmetry invariant case. AB caging has been first introduced in a two-dimensional Dice lattice structure.\cite{vidal1998aharonov} From the first case, the study of this localization has been further extended~\cite{doucot2002pairing,fang2012photonic,longhi2014aharonov,kibis2015aharonov,hasan2016optically,ZhangJin_PRB_102_054301_2020} while experimentally AB caged systems has been realized in photonic lattices~\cite{fang2012realizing,mukherjee2018experimental} and qubits nanocircuits~\cite{gladchenko2009superconducting}.

Since an ABF model has all bands flat, the system displays time reversal invariance. Therefore there is no need to search for ABF models by detouring via time-reversal broken
systems with nonzero magnetic synthetic fluxes or simply magnetic fields, only to re-arrive back to a time-reversal symmetry invariant case. 
Instead, 
it is much simpler to finetune time-reversal symmetry invariant tight-binding manifolds in order to isolate the ABF cases. This task was executed
in Ref.~\onlinecite{danieli2020cagingI} through a set of local unitary detangling transformations and their inverse entangling ones, leading to a generator
of ABF models in any lattice dimension with any number of flatbands. The detangled basis is the preferred one for the study of perturbations such as
disorder and interactions.

The impact of interactions on caged particles was studied in the 1D Creutz lattice~\cite{creutz1999end,takayoshi2013phase,tovmasyan2013geometry,tovmasyan2016effective,junemann2017exploring,tovmasyan2018preformed} and the 2D Dice lattice~\cite{vidal2001disorder}. In both these cases as well as in the 1D AB diamond (rhombic) lattice it has been shown that the Hubbard interaction breaks the single particle caging for some eigenstates by inducing transporting bound states of paired particles.~\cite{vidal2000interaction,tovmasyan2018preformed} 
Transporting bound states co-exist with the trivial two particle states with particles caged in two CLS which are separated beyong the reach of the 
Hubbard interaction and which therefore remain exact eigenstates of the two particle system. 
Remarkably, some macroscopically degenerate eigenstates continue to lack dispersion but show energy renormalization upon tuning the Hubbard interaction strength~\cite{vidal2000interaction}.

In a related work in Ref.~\onlinecite{danieli2020cagingI} we introduced an ABF generator scheme in any lattice dimension. 
Each set of ABF models is related to one detangled ABF lattice. For a given member of the set, additional
short-range nonlinear interactions destroy caging in general and induce transport. However, fine tuned subsets allow to completely restore caging.
We derived necessary and sufficient fine-tuning conditions for nonlinear caging including computational evidence.

Here we replace nonlinear terms by Bose-Hubbard interactions and study quantum caging.
We prove the existence of degenerate energy renormalized compact states for two and three particles, and inductively conjecture to generalize to any finite number M of participating particles in one dimension.
Our results explain and generalize
previous observations for two particles on a diamond chain~\cite{vidal2000interaction}.
We further prove that quantum caging conditions  guarantee the existence of extensive sets of conserved local parities in any lattice dimension, as first revealed in Ref. \onlinecite{tovmasyan2018preformed}
for a set of specific one-dimensional networks. 
Consequently transport is realized through moving pairs of interacting particles which break the single particle caging.

\section{Single particle detangling and entangling: ABF generator scheme revisited}

In a related work in Ref.~\onlinecite{danieli2020cagingI} it was shown that a single particle ABF network can be de- and entangled with local
unitary transformations, arriving at a generator of ABF networks. Assuming a translationally invariant tight binding network, a local unitary
transformation set $U_1$ is given by a product of commuting local unitary transformations (LUT). Each LUT acts on a spatially localized
Hilbert subspace, and all LUTs can be obtained from one by all possible discrete space translations. Given a 1D ABF network with short range
hopping, it was shown~\cite{danieli2020cagingI}  that a finite number of non-commuting LUT sets $U_1,...,U_u$ will detangle the Hamiltonian
into a completely diagonal form. The number $(u+1)$ corresponds to the hopping range. Inverting this procedure leads to the most general and exhausting
ABF generator with the elements of the LUTs being the relevant control parameters.   Extensions to higher lattice dimensions $d=2,3,...$ 
result in ABF generators with the number of unitary sets increasing $\sim d(u+1)$.

With $\nu$ flatbands, the detangled parent Hamiltonian is diagonal with 
\begin{equation}
\hat{\mh}_\text{fd} = \sum_k \hat{f}_k
\label{eq:H_fd}
\end{equation}
with each local Hamiltonian $\hat{f}_k$ being diagonal and of rank $\nu$. All local Hamiltonians have identical eigenvalue sets. 
The corresponding eigenvectors are compact localized states in real space. The first entangling step consists of applying one
LUT $U_1$. The resulting family of ABF Hamiltonians is coined {\sl semi-detangled} (SD) and is a sum over commuting local Hamiltonians~\cite{danieli2020many}
\begin{equation}
\hat{\mh}_\text{sd} = \sum_k \hat{s}_k
\label{eq:H_sd}
\end{equation}
with each $\hat{s}_k$ of rank $\nu$. 
The system still shows trivial compact localization - dynamics is restricted to each of the $\nu$-mers related to a specific $\hat{s}_k$.
Adding $d$ (non-commuting) LUTs $U_2 ...U_{d+1}$ results in a nontrivial entangling and full connectivity on the entire network.
The choice of the LUTs and their matrix elements fix one particular ABF family member.

\section{Many body interactions}
\label{sec:MBI}

We choose a particular ABF family member and add onsite Bose-Hubbard interaction
\begin{gather}
    \hmhb_1 = \frac{U}{2}\sum_{n\in\mathbb{Z},a} \hc_{na}^{\dagger} \hc_{na}^{\dagger}\hc_{na}\hc_{na},
    \label{eq:Hubbard}
\end{gather}
where $\hc_{n,a})$ denote bosonic annihilation operators in a unit cell $n$ with $1 \leq a \leq \nu$. The replacement of the operators by c-numbers
was considered in Ref.~\onlinecite{danieli2020cagingI} which addressed the question whether there is an ABF family submanifold which preserves
nonlinear caging. The submanifolds were derived by transforming an ABF family member back into its detangled parent basis. 

For the quantum counterpart, such a transformation turns 
the above interaction into
\begin{gather}
    \hmhb_1 = \frac{U}{2}\sum_{na,mb;kc,ld} V_{na,mb;kc,ld} \hc_{n,a}^\dagger\hc_{m,b}^\dagger\hc_{k,c}\hc_{l,d}
    \label{eq:Hint_nd}
\end{gather}
in the detangled representation. These terms represent generic density assisted hopping terms for pairs of particles, implying that without any fine-tuning, either of the interaction or the single particle Hamiltonian, the caging is broken and  there should be transport in the interacting problem.

This is in accordance with the previous studies that indicate that we should expect emergence of transporting states: Vidal \textit{et. al}~\cite{vidal2000interaction} predicted extended states in the Aharonov-Bohm diamond chain with Hubbard interaction already for two particles. Tovmasyan \textit{et al.}~\cite{tovmasyan2018preformed} confirmed that and furthermore conjectured the existence of an extensive sets of conserved quantities -- number parity operators -- to be a generic feature of interacting ABF networks based on the several models studied.

We now provide a proof of this conjecture for the necessary and sufficient conditions on the interactions for the existence of extensive set of conserved quantities. We also discuss potential generalisations of this result. The proof follows very naturally from our results for classical nonlinear interactions: conserved quantities -- number parity operators -- appear every time the interacting AFB network only allows particles to move in pairs between the unit cells, or, equivalently whenever the classical version of the network features caging. Indeed as we showed in Ref.~\onlinecite{danieli2020cagingI}, nonlinear caging occurs in classical models \emph{iff} the interaction, in the detangled basis, takes the following form (for the Kerr-like nonlinearity, which corresponds to the Hubbard-like interaction in the quantum case):
\begin{align}
    \label{eq:gen-ham-caged}
    \mh_1 = & \sum_{na,b;mc,d} V_{na,nb;mc,md} \phi_{n,a}^*\phi_{n,b}^*\phi_{m,c}\phi_{m,d}\\
    + & \sum_{na,b;mc,d} V_{na,mb;nc,md} \phi_{n,a}^*\phi_{m,b}^*\phi_{n,c}\phi_{m,d}, \notag
\end{align}
where $\phi_{n,a}$ is a classical amplitude on the site $a$ inside unit cell $n$. The exact choice of the interaction is not relevant for the proof. The quantum version of this Hamiltonian in the second quantised form reads:
\begin{align}
    \label{eq:gen-ham-caged}
    \hmhb_1 = & \sum_{na,b;mc,d} V_{na,nb;mc,md} \hc_{n,a}^\dagger\hc_{n,b}^\dagger\hc_{m,c}\hc_{m,d}\\
    + & \sum_{na,b;mc,d} V_{na,mb;nc,md} \hc_{n,a}^\dagger\hc_{m,b}^\dagger\hc_{n,c}\hc_{m,d}, \notag
\end{align}
Such interaction, which we loosely term a {\sl quantum caging} one, only allows particles to hop in pairs between the unit cells. The terms in the above two sums have immediate interpretation: the first sum only contains terms that allow pairs of particles to hop from one unit cell to another, while the terms in the second sum simply swap particles between two different unit cells. In other words, the terms in the first sum can only change the number of particles in a unit cell by $2$, while the terms in the other sum cannot change this number. The full Hamiltonian therefore commutes with the parity operator of the number of particles in each unit cell (remember that single particle Hamiltonian does not move particles at all). This proves the sufficiency of the condition, while the necessity is self-evident. 
There can be caged isolated particles in the system, since for an odd number of particles in a unit cell, one particle is doomed to stay in that unit cell forever, since the interaction is unable to move it~\cite{tovmasyan2018preformed}. Note also that a system of spinless fermions freezes completely for 
a quantum caging interaction, since double occupancy is forbidden.

Are there additional caging features in these systems not captured by such operators? 
Such a question appears reasonable following the prediction of non-dispersive states of two interacting spinful fermions in the Aharonov-Bohm diamond chain~\cite{vidal2000interaction} -- states which escape the counting argument for single caged particle above.
Note that the cases considered in Ref.~\onlinecite{tovmasyan2018preformed} correspond to a fine-tuning of the 1D models where particles have to move in pairs between the unit cells (in the detangled representation).
The observation of similar features for the 2D Dice lattice~\cite{tovmasyan2018preformed} provides an indirect evidence in support of our conjecture that ABF lattices in any dimension can be detangled.~\cite{danieli2020cagingI}

We note that it is also possible to perform a further fine-tuning of the interaction and eliminate the pair interaction assisted hopping, and turn the second sum into a pure density-density interaction, leading to no particle transport, e.g. a perfect charge insulator.~\cite{tovmasyan2018preformed,danieli2020many,kuno2020flat,orito2020exact}
Finally, just like in the classical case, we can use this construction to test whether a given combination of an ABF network and many-body interaction 
possesses conserved quantities -- by inspecting the interaction in the detangled basis. We can also invert the procedure and start with a given interaction that moves particles in pairs in the detangled basis, and then apply a unitary transformation and get into some other basis, resulting in general in a model with complicated many-body interactions.

\section{Two bands networks}

We consider interacting bosons evolving on the $\nu=2$ ABF networks introduced in~\cite{danieli2020cagingI} and governed by the Bose-Hubbard Hamiltonian $\hmh = \hmhb_0 + \hmhb_1$ with
\begin{align}
    \label{eq:BH1}
    \hmhb_0 &= - \sum_{n\in\mathbb{Z}} \left[\frac{1}{2} \left( \hat{c}_{n}^{+ T} H_0\hat{c}_n \right) +  \left(\hat{c}_{n}^{+ T} H_1\hat{c}_{n+1} \right) + h.c.\right],\\
    \hmhb_1 &= \frac{U}{2}\sum_{n\in\mathbb{Z}}\left[ \ha_{n}^{\dagger} \ha_{n}^{\dagger}\ha_{n}\ha_{n} + \hb_{n}^{\dagger} \hb_{n}^{\dagger} \hb_{n} \hb_{n} \right].
    \label{eq:BH1_TBI}
\end{align}
Here $n$ labels the unit cell; the annihilation $\hc_n = (\ha_n, \hb_n)$ and creation operators $\hc_n^\dagger = (\ha_n^\dagger, \hb_n^\dagger)$ respect the commutation relations $[\ha_n, \ha_k^\dagger] = \delta_{n,k}$, $[\hb_n, \hb_k^\dagger] = \delta_{n,k}$, and $[\ha_n, \hb_k^\dagger] = 0$ for any $n,k\in\mathbb{Z}$. The hopping matrices
\begin{align}
    \label{eq:H0H1_rot0}
    H_0 &= \Gamma_0
    \begin{pmatrix}
        |z_1|^2  -  |w_1|^2 &  -2  z_1 w_1 \\[0.3em]
        -2 z_1^* w_1^* &   |w_1|^2 - |z_1|^2
    \end{pmatrix},\\
    & H_1 = \Gamma_1
    \begin{pmatrix}
        z_1  w_1^* &  z_1^2  \\[0.3em]
        - (w_1^*)^2 &- z_1  w_1^*
    \end{pmatrix}
    \label{eq:H0H1_rot1}
\end{align}
are hopping inside a unit cell and between the n.n. unit cells respectively; $\Gamma_0 =  |w_2|^2 -  |z_2|^2$ and $\Gamma_1 = 2 z_2 w_2$. Here $z_i,w_i$ are complex numbers such that $|w_i|^2 +  |z_i|^2=1$ for $i=1,2$, and the single particle Hamiltonian $\hmhb_0$ has two flatbands at $E=\pm 1$ -- see Ref.~\onlinecite{danieli2020cagingI}.

\subsection{Breakdown of caging and conserved quantities}
\label{sec:2IP_breaking}

Let us first illustrate the above general results by comparing the quantum case to the classical one. 
To do so, it is sufficient to recast $\hmhb_0$ in its semi-detangled representation $\hat{\mh}_\text{sd}$~\eqref{eq:H_sd} rather than its detangled parent $\hat{\mh}_\text{fd}$~\eqref{eq:H_fd}. 
This turns the Hubbard interaction $\hmh_1$ in Eq.~\eqref{eq:BH1_TBI} into a nonlocal interaction Eq.~\eqref{eq:Hint_nd} -- the full expression can be found in Appendix~\ref{app3b}. 
If the nonlinear caging fine-tuning $|z_1|^2 = |w_1|^2$ is satisfied, then for the annihilation and creation operators $\hp_{n}, \hf_{n}$ and $\hp_{n}^{\dagger},\hf_{n}^{\dagger}$, the interaction $\hmh_1$ reads
\begin{align}
    \hmhb_1 & = U\sum_n \left\{ |z_1|^4 \left[ \hp_n^\dagger\hp_n^\dagger \hp_n\hp_n + \hf_n^\dagger\hf_n^\dagger \hf_n\hf_n + 4 \hp_n^\dagger\hp_n \hf_{n+1}^\dagger\hf_{n+1}\right] \right. \notag\\
    \label{eq:Ham1_BH_R}
    & \left. \qquad\quad + z_1^{*2} w_1^2 \hp_{n}^\dagger \hp_{n}^\dagger \hf_{n+1}\hf_{n+1} + z_1 w_1^{*2} \hp_{n} \hp_{n} \hf_{n+1}^\dagger \hf_{n+1}^\dagger \right\}.
\end{align}

\begin{figure}
    \centering
    \includegraphics[width=0.95\columnwidth]{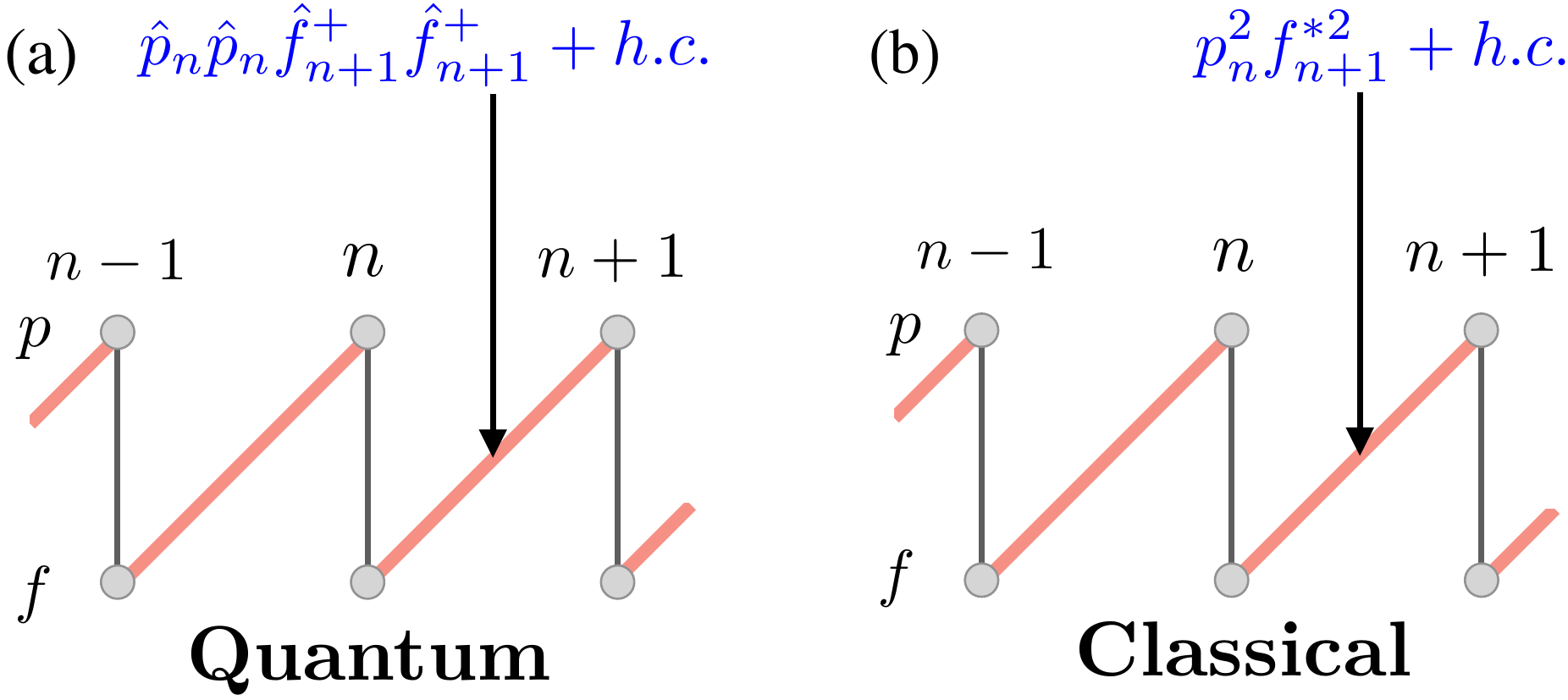}
    \caption{Schematic representation of linear hopping (solid gray lines) and interaction terms (red shaded lines) for the  quantum Hubbard case (a) and the classical nonlinear case (b).}
    \label{fig:interaction}
\end{figure}

The terms in the first line of Eq.~\eqref{eq:Ham1_BH_R} are products of particle number operators -- hence, they do not move any particles from site to site. Let us focus on the terms in the second line of Eq.~\eqref{eq:Ham1_BH_R} along with their classical counterparts -- shown with red shaded lines in Fig.~\ref{fig:interaction} (a) and (b) respectively. 
Both quantum and classical terms connect the decoupled single particle dimers (solid gray lines in Fig.~\ref{fig:interaction}). 
The quantum terms $\hp_n \hp_n \hf_{n+1}^\dagger\hf_{n+1}^\dagger +$ h.c. apply as soon as at least two particles access any of the two sites -- irrespective of the presence of particles on the other site --  and they provide coherent transport for pairs of particles along the ladder, in agreement with the number parity operators introduced in Ref.~\onlinecite{tovmasyan2018preformed} and discussed in Sec.~\ref{sec:MBI}.
The corresponding classical terms $p_n^2 f_{n+1}^{* 2} +$ h.c. turn nonzero  if the amplitudes on both sites in $p_n$ and $f_{n+1}$ are nonzero -- therefore preserving nonlinear caging if one of them vanishes.

\subsection{Dynamics of two interacting particles}
\label{sec:qc}

Let us propagate two interacting particles initially located at the same site -- hence avoiding unpaired caged particles.~\cite{tovmasyan2018preformed} 
For convenience and without much loss of generality, we consider distinguishable bosons. We compute the time-evolution of their wave-function $\ket{\Psi}$  which is governed by a two-dimensional lattice Schr\"odinger equation whose coordinates $(n,k)$ represent the spatial position of each boson -- see Appendix~\ref{app:Mdim_sch}. We consider a network for $N=40$ unit cells and compute the local density $\rho_{n,k}$ of the two particles and the correspondent one-dimensional probability distribution function (PDF) of the particle density defined as $Q_n = \sum_{k=1}^N  \rho_{n,k}$. 

As the single particle Hamiltonian $\hmhb_0$ testbed, we consider the quantum version of \emph{model A}~\cite{danieli2020cagingI}
\begin{align}
    \hmhb_0  = \sum_n & \left[- 2 \ha_n^\dagger\hb_{n}+  \sqrt{3}\left(\ha_n^\dagger\ha_{n+1} + \ha_n^\dagger\hb_{n+1} \right.\right. \notag \\
    \label{eq:modelA-H0}
    &\qquad\qquad\quad \left.  \left. - \hb_n^\dagger\ha_{n+1}  -\hb_n^\dagger\hb_{n+1} \right) + \text{h.c.}\right]
\end{align}
whose classical version satisfies the nonlinear caging condition  $|z_1|^2 = |w_1|^2$ in the presence of Kerr nonlinearity, 
and \emph{model B}~\cite{danieli2020cagingI}
\begin{align}
    \hmhb_0  = \sum_n & \left[\sqrt{3} \ha_n^\dagger\ha_{n+1}  + 3 \ha_n^\dagger\hb_{n+1} \right. \notag \\  
    & \left. - \ha_n^\dagger\hb_{n-1}  -\sqrt{3} \hb_n^\dagger\hb_{n+1}   + \text{h.c.}\right]
    \label{eq:modelB-H0}
\end{align}
whose classical version breaks the caging condition for the same nonlinearity. 
Note that both Hamiltonians are related to each other and to their fully detangled parent ABF Hamiltonian through proper LUTs.
We now add the Bose-Hubbard interaction (\ref{eq:BH1_TBI}) to both models and evolve the two particles.

\begin{figure}
    \centering
    \includegraphics[width=0.95\columnwidth]{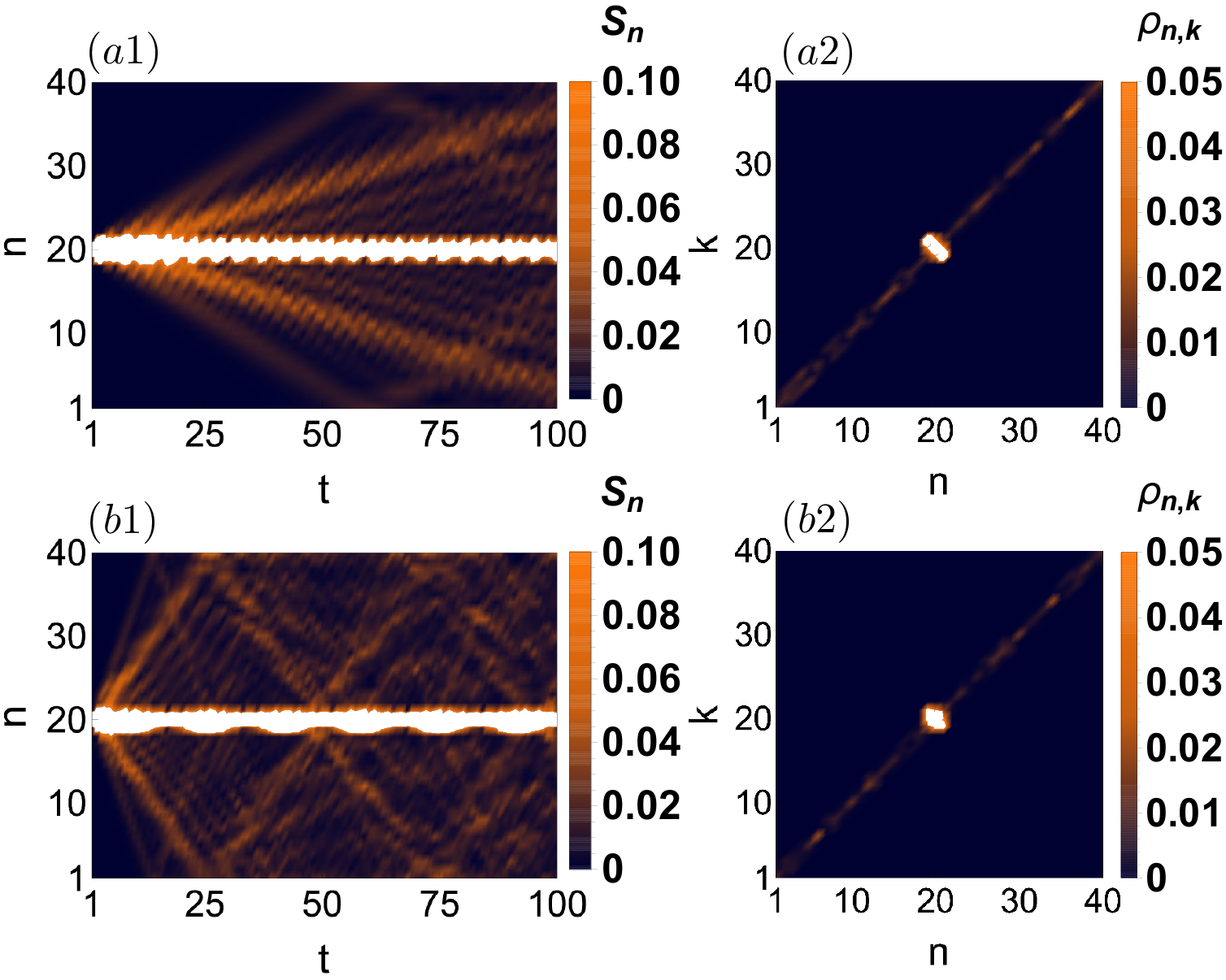}
    \caption{Model A, Eq.~\eqref{eq:modelA-H0} combined with the Hubbard interaction satisfies the condition for the existence of conserved quantities. (a1) Time evolution of $Q_n$ and (b1) local density $\rho_{n,k}$ at time $t=100$, both for $U=1$. (a2-b2) Same as (a1-b1) but for $U=5$.}
    \label{fig:ex1_2IP}
\end{figure}

In Fig.~\ref{fig:ex1_2IP} (a1-b1) we plot the time-evolution of $Q_n$ for \emph{model A} for two interaction strengths $U=1$ (top) and $U=5$ (bottom) respectively. Both cases show that a part of the PDF is propagating ballistically - indicating the spreading of the particle pair along the network, breaking single particle caging. Simultaneously, we observe that a substantial portion of $Q_n$ remains localized at unit cell $n=\frac{N}{2}$ (the initial location of both particles). This is further detailed in Fig.~\ref{fig:ex1_2IP} (a2-b2), where we plot the local density $\rho_{n,k}$ at time $t=100$ for $U=1$ and $U=5$ respectively. Firstly, these panels show that the delocalization and ballistic spreading of a part of the wave-function $\ket{\Psi}$ occurs along the diagonal $k=n$, which indicates that the particles have to stay bound in order to delocalize. Secondly, these plots show a large amplitude density peak at the original launching site. Such long-lasting localized excitations hint at the existence of non-propagating spatially compact states of two interacting particles which have been excited by placing both particles initially in the same cell -- akin to those predicted in the  AB diamond chain.~\cite{vidal2000interaction}

\begin{figure}
    \centering
    \includegraphics[width=0.95\columnwidth]{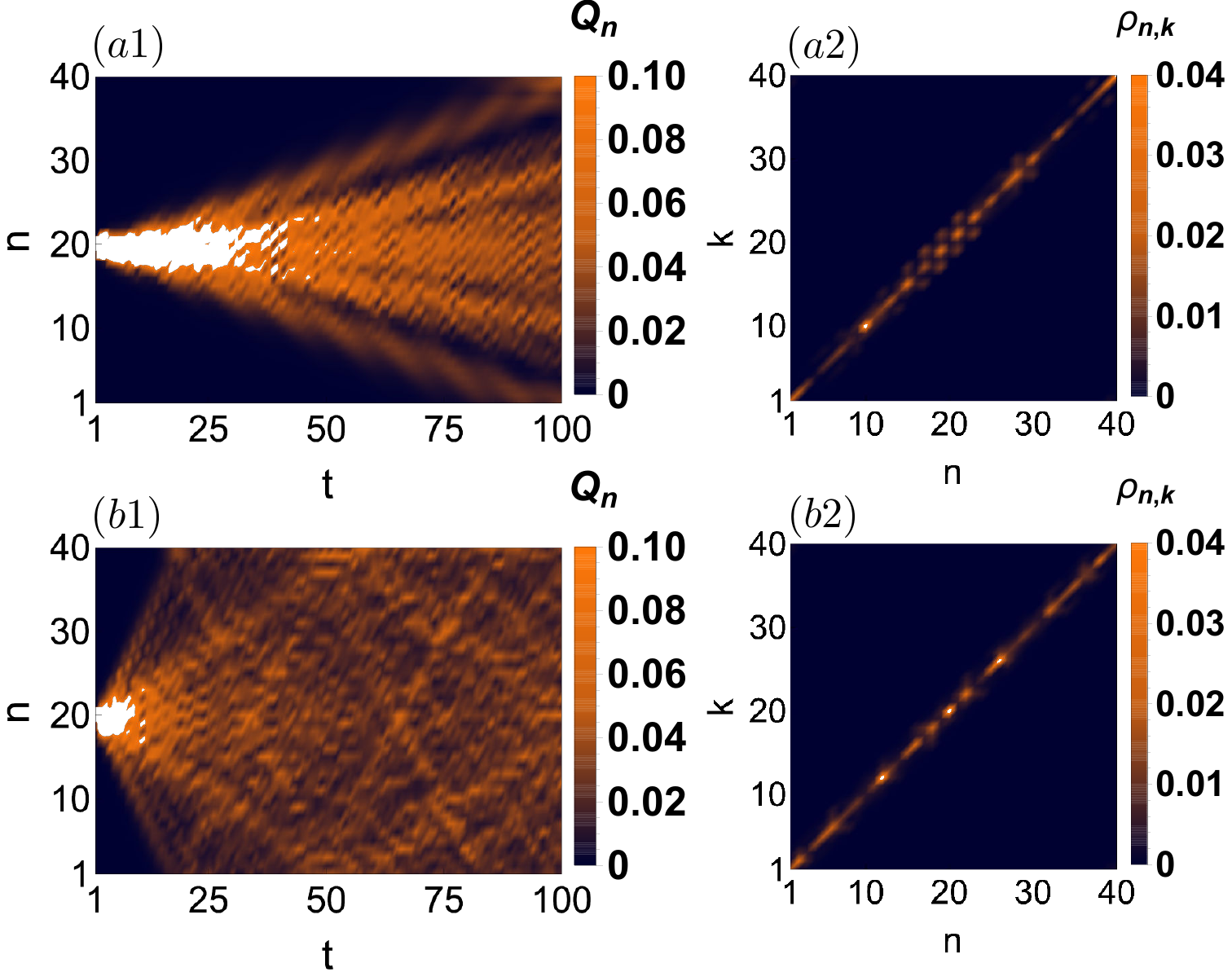}
    \caption{Model B, Eq.~\eqref{eq:modelA-H0}, combined with the Hubbard interaction does not possess conserved quantities. (a1) Time evolution of $Q_n$ and (b1) local density $\rho_{n,k}$ at time $t=100$, both for $U=1$. (a2-b2) Same as (a1-b1) but for $U=5$.}
    \label{fig:ex2_2IP}
\end{figure} 
 
As expected, the signatures of caged states are completely absent for \emph{model B} that does not obey the fine-tuning condition. Indeed, in Fig.~\ref{fig:ex2_2IP} (a1-b1) we show the time-evolution of $Q_n$ for $U=1$ (top) and $U=5$ (bottom) respectively. These plots again show ballistic spreading, i.e. the existence of spatially extended bound states. However, no considerable localized fraction of the PDF $Q_n$ is observed.
This is also confirmed in Fig.~\ref{fig:ex2_2IP} (a2-b2), where the the local density $\rho_{n,k}$ at time $t=100$ is shown for $U=1$ and $U=5$ respectively.

\subsection{Interaction renormalized compact states}

To explain and generalize the results in Fig.~\ref{fig:ex1_2IP} for model A, we consider a finite number $M$ of interacting indistinguishable bosons.
The wave function $\ket{\psi}$ of $M$ interacting particles can be expanded in a Fock states basis of $M$ particles as 
$\ket{\psi} = \sum_{{\mathbf n}} \varphi_{\mathbf n} \cdot \ket{v_{\mathbf n}}_{\bf a,b}$ -- see Appendix~\ref{app:Mdim_sch}.
Its time evolution is governed by the M-dimensional lattice Schr\"odinger equation 
\begin{gather}
    i \dot{\varphi}_{\mathbf n} = \left[A + U V_M \right]\varphi_{\mathbf n} + \sum_{j=1}^M \left[ T_j \varphi_{{\mathbf n} + {\mathbf e}_j}  + T_j^\dagger \varphi_{{\mathbf n} - {\mathbf e}_j}\right].
    \label{eq:Mip_int_eqs}
\end{gather}
Here $\varphi_{\mathbf n}$ is a complex vector with $2^M$ components, while $A$ and $\{T_j \}_{j\leq M}$ are square matrices describing the dynamics of the noninteracting particles. The diagonal matrix $V_M$ encodes the interaction between particles described by the Hamiltonian $\hmhb_1$~\eqref{eq:BH1_TBI}.

The Hamiltonian $\hmhb_0$ describes $M$ non-interacting particles and can be fully detangled via a sequence of unitary transformations.~\cite{danieli2020cagingI} Likewise, the $M$-dimensional Schr\"odinger system Eq.~\eqref{eq:Mip_int_eqs} for $U=0$ has only flatbands, and it can be mapped to a fully disconnected network (Appendix~\ref{app:detangling_M}). We now apply the detangling procedure to Eq.~\eqref{eq:Mip_int_eqs} in the presence of the interaction $U\neq0$. The matrix $V_M$ induces a coupling network between the detangled sites which describes dispersive states of interacting particles and -- under the fine-tuning condition $|z_1|^2 = |w_1|^2$ highlighted in Ref.~\onlinecite{danieli2020cagingI} -- renormalized compact states of interacting particles. 
We will first consider $M=2$ and apply induction for $3\leq M$.

\subsubsection{Two particles} 
\label{sec:2IP}

\begin{figure}
    \centering 
    \includegraphics[width=0.925\columnwidth]{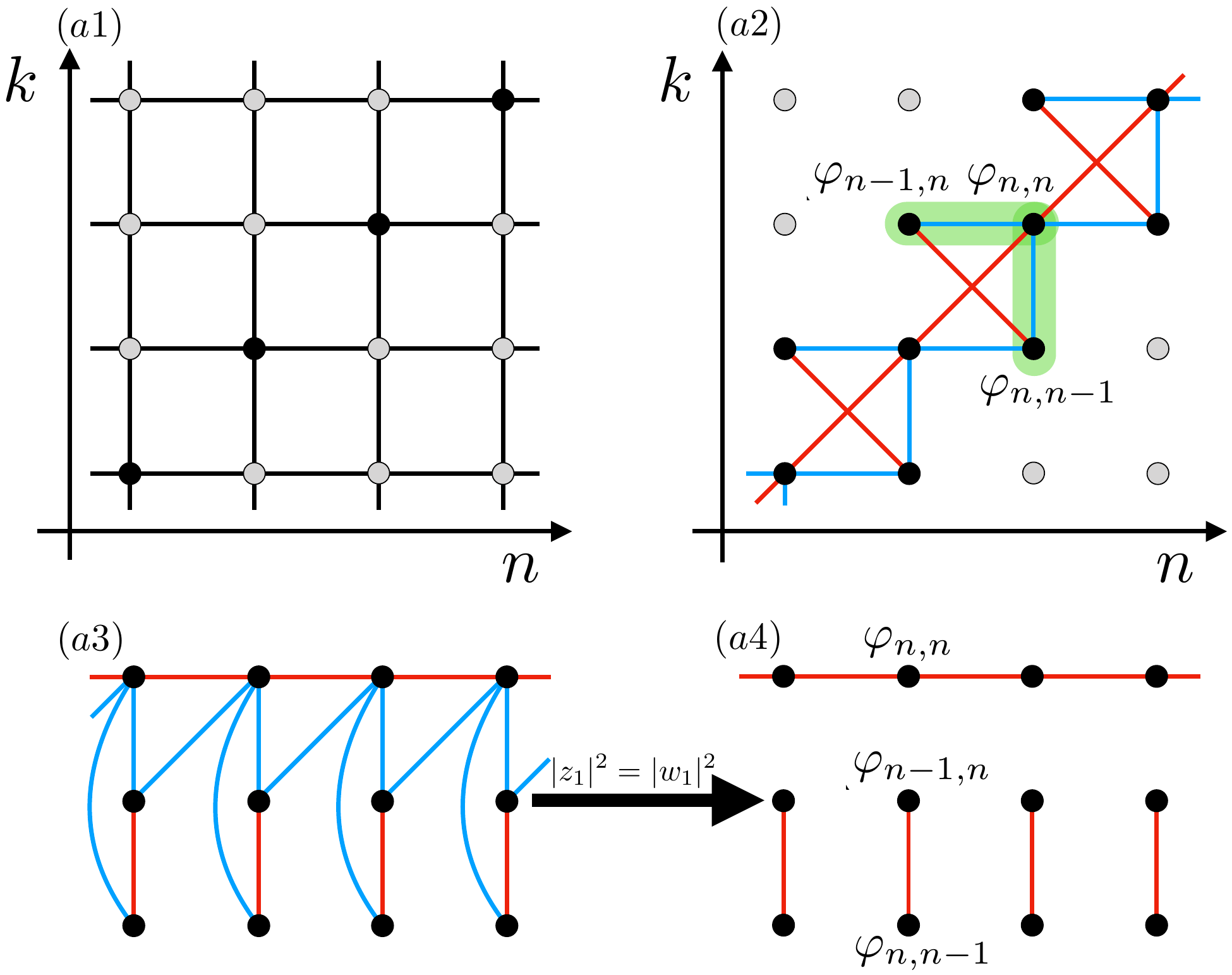}
    \caption{(a1) 2D network in Eq.~\eqref{eq:Mip_int_eqs} for $M=2$ particles. Each dot indicates one unit cell $\varphi_{n,k}$, and the black lines represent the hopping between cells. The black dots indicate the main diagonal $k=n$ where interaction applies. (a2) Rotated network. (a3) Unfolding into a 1D chain in Eq.~\eqref{eq:network_diag} along the main diagonals  of the rotated 2D network in (a2).
(a4) Same as (a3) for $|z_1|^2 = |w_1|^2$. The blue links in (a3) disappeared leaving components detangled from the dispersive chain and representing the renormalized compact states. }
    \label{fig:2D_mapped}
\end{figure}

For $M=2$ interacting particles, Eq.~\eqref{eq:Mip_int_eqs} reduces to a two dimensional network -- shown in Fig.~\ref{fig:2D_mapped}(a1).  
In the noninteracting case, this network has four flatbands at $E=\pm2$ and $E=0$ (doubly degenerate) and it can be mapped to a fully disconnected 2D lattice.  
In this new representation, the interaction matrix $V_2$ induces a coupled network along the diagonals $k=n,n\pm 1$ of Eq.~\eqref{eq:Mip_int_eqs}, governed by the equation
\begin{gather}
    i \dot{\eta}_n = K_D \eta_n + H_D \eta_{n+1} + H_D^\dagger \eta_{n+1}
    \label{eq:network_diag}
\end{gather} 
for the unit cell $\eta_n = (\varphi_{n-1,n} , \varphi_{n,n} , \varphi_{n,n-1})$ -- as shown in Fig.~\ref{fig:2D_mapped}(a2).

For $k\neq n,n\pm 1$ -- away from the diagonal -- the rotated network Eq.~\eqref{eq:Mip_int_eqs} remains fully detangled, since these sites correspond to product states of two particles caged far from each other  and therefore insensitive to the Hubbard interaction. At the diagonal lines $k=n,n\pm 1$, the 1D chain Eq.~\eqref{eq:network_diag} - highlighted in Fig.~\ref{fig:2D_mapped}(a3) - yields transporting extended states representing two interacting particles coherently evolving along the system and breaking the single particle caging -- as was discussed in Sec.~\ref{sec:qc}. However, the fine-tuning $|z_1|^2 = |w_1|^2$ reduces Eq.~\eqref{eq:network_diag} to a lattice with detangled compact interacting states similar to previously reported detanglings of CLSs
in the presence of other dispersive states~\cite{flach2014detangling} (see Fig.~\ref{fig:2D_mapped}(a4) and details in Appendix~\ref{app5_fano}). This implies that:
\begin{itemize}
    \item[(i)] the diagonal $k=n$ in the $\varphi_{n,n}$ coordinates yields dispersive states of two paired particles freely moving along the chain -- see Ref.~\onlinecite{tovmasyan2018preformed} and Sec.~\ref{sec:MBI};
    \item[(ii)] the detangled component described by $(\varphi_{n-1,n} , \varphi_{n,n-1})$ yields degenerate compact states of two interacting particles with renormalized $U$-dependent energies.
\end{itemize}
Let us compute explicitly the energies of both types of states -- dispersive and compact ones -- for 
an example of a $\nu=2$ lattice satisfying  $|z_1|^2 = |w_1|^2$. 
Without loss of generality, to simplify our analytics instead of choosing model A in Eq.(\ref{eq:modelA-H0}) we present our results for a test case obtained for $z_i = \cos \pi/4 $, $w_i = \sin \pi/4$ for $i=1,2$ in Eqs.~(\ref{eq:H0H1_rot0},\ref{eq:H0H1_rot1}).
The dispersive component in $\varphi_{n,n}$ leads to a characteristic polynomial (Appendix~\ref{app5_CLS_ex1})
\begin{align}
    p_D(E,k;U) & = E  [E^3 - U E^2  -16 E + 8 U(1-\cos k) ] \notag\\
    & \equiv E \, g_D(E,k;U) 
    \label{eq:2p_int_disp_charpoly}
\end{align}
yielding four bands: one flat at $E=0$ and three dispersive bands given by zeroes of $ g_D(E,k;U)$. The decoupled component $(\varphi_{n-1,n} , \varphi_{n,n-1})$ leads to the characteristic polynomial (Appendix~\ref{app5_CLS_ex1})
\begin{align}
    p_F(E;U) & = E^3 ( E^2 -16 ) (E^3 - U E^2  -16 E + 8 U ) \notag \\
    & \equiv E^3 ( E^2 -16 ) \cdot g_F(E;U)
    \label{eq:2p_int_offdiag_charpoly}
\end{align}
yielding five non-renormalized bands at $E=0,\pm4$ and three bands with $U$-renormalized energies as zeroes of $g_F(E;U)$. Notably,  Eqs.~(\ref{eq:2p_int_disp_charpoly},\ref{eq:2p_int_offdiag_charpoly}) are very similar to those obtained for two interacting spinful fermions AB diamond chain~\cite{vidal2000interaction}.

\begin{figure}
    \centering 
    \includegraphics[width=0.95\columnwidth]{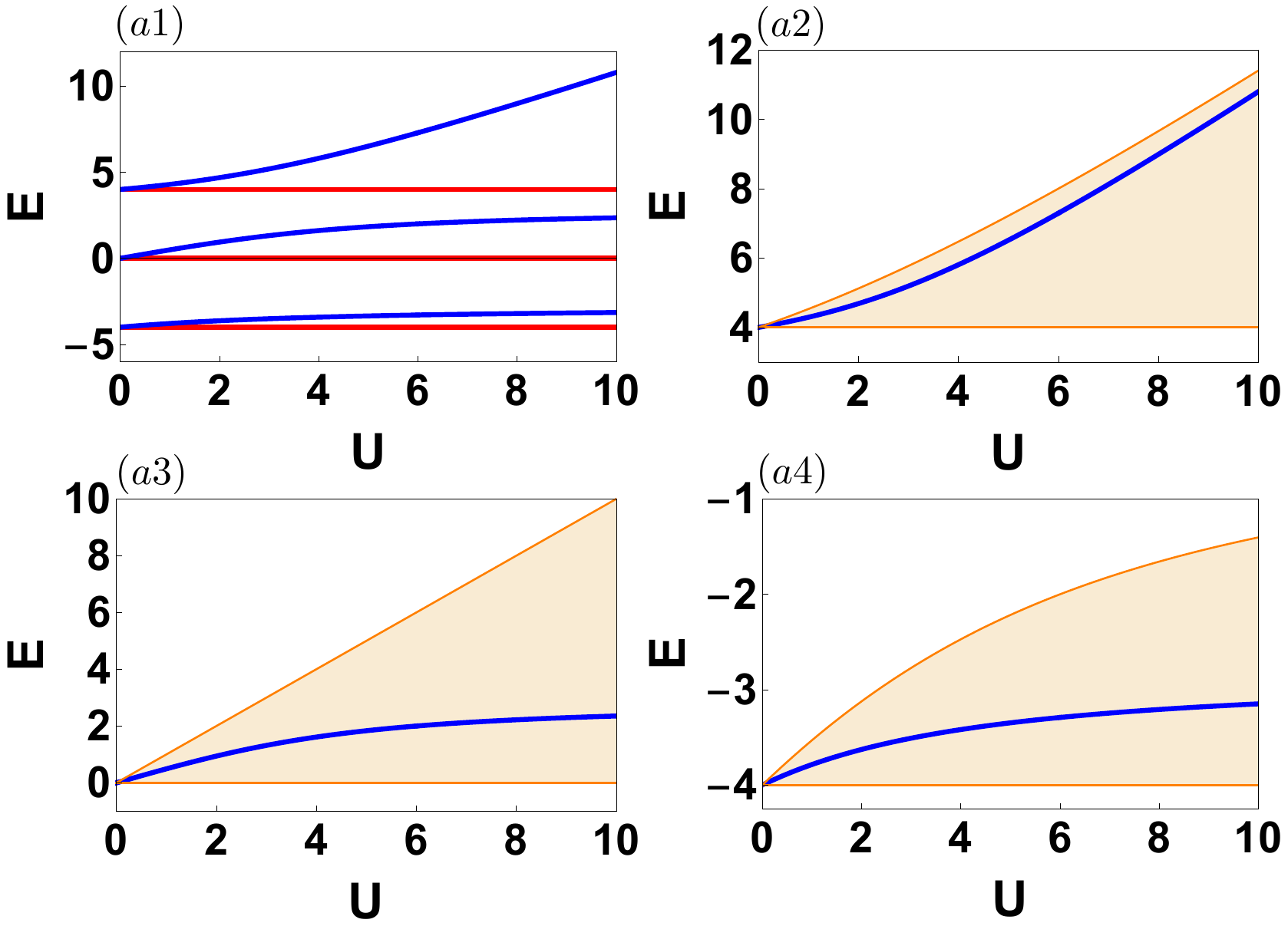}
    \caption{(a1) Non-renormalized degenerate energies (red lines) of two noninteracting CLSs and renormalized energies (blue curves) of two 
compactly bound interacting particles versus $U$. (a2-a4) Renormalized energies (blue curve) within the dispersive bands (orange areas) of 2IP bound states.
}
    \label{fig:ex1_2IP_energies}
\end{figure}

In Fig.~\ref{fig:ex1_2IP_energies}(a1) we show the non-renormalized (red horizontal lines) and the renormalized degeneracies (blue curves) versus the interaction strength $U$. In the remaining three panels Fig.~\ref{fig:ex1_2IP_energies}(a2-a4) we plot each renormalized degeneracy shown in panel (a1) and one of the three dispersive bands (orange shaded areas) obtained as zeroes of $g_D(E,k;U)$ in Eq.~\eqref{eq:2p_int_disp_charpoly}. 
We also observe that all the renormalized energies (blue curves) lie within one of the dispersive bands -- as for any $U$ it holds that $g_F(E,U) = g_D(E,0;U)$ -- characterizing these renormalized compact states as quantum  two particles bound states in the continuum (BIC).~\cite{hsu2016bound}

\begin{figure}
    \centering
    \includegraphics[width=\columnwidth]{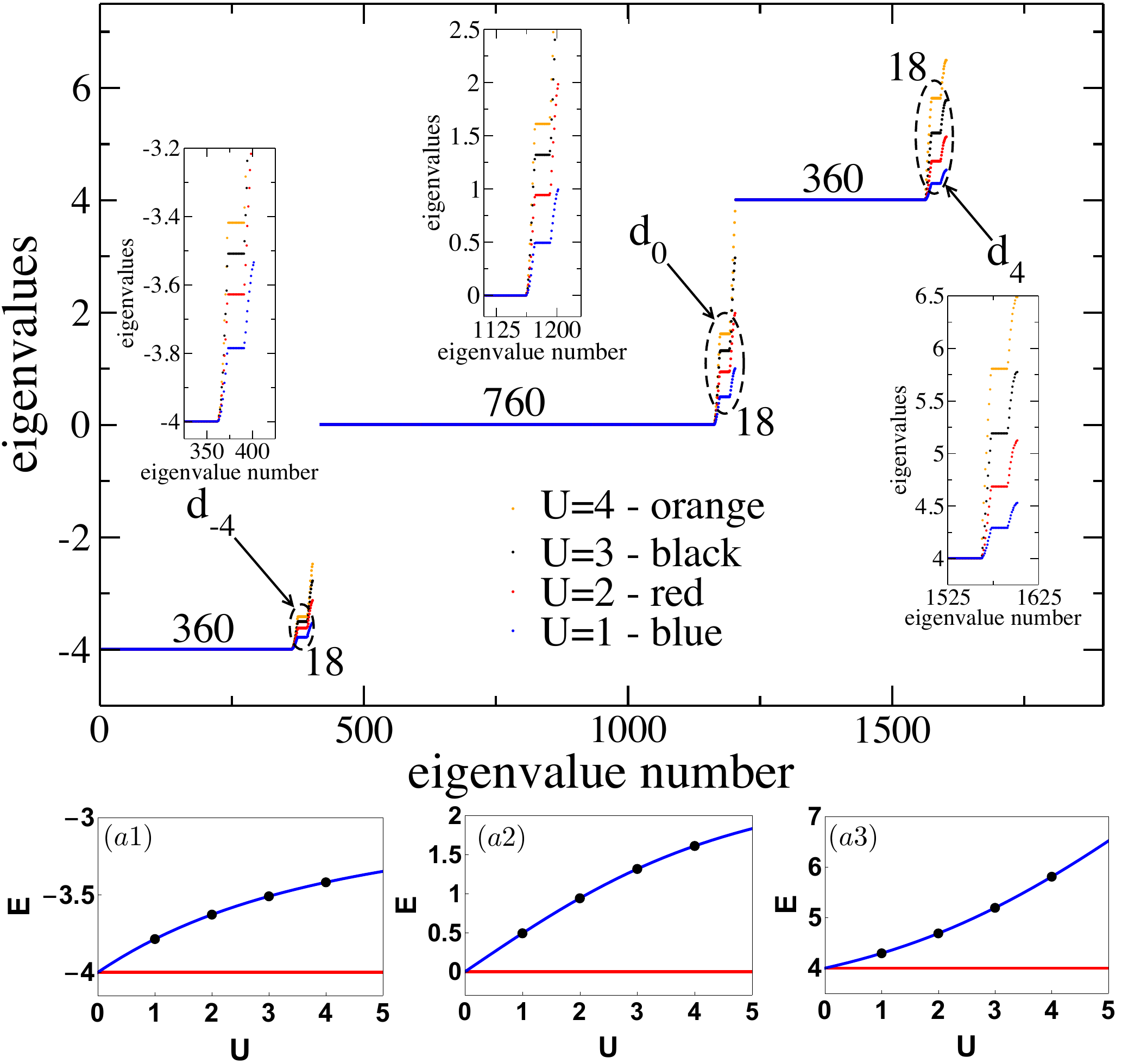}
    \caption{(main) Eigenvalues of $M=2$ interacting bosons for $N=20$. The numbers count the degenerate eigenvalues.   
    (a1-a3) Renormalized energies (blue curves) versus $U$ and numerically obtained degeneracies (black dots) at the region $d_{-4}$ (a1),  $d_0$ (a2), and  $d_4$ (a3).}
    \label{fig:ex1_2IP_energies_num}
\end{figure}

We tested these results numerically by diagonalizing the two dimensional network given by Eq.~\eqref{eq:Mip_int_eqs} for $M=2$ bosons at $U=1,2,3,4$ and $N=20$ unit cells. The results are reported in the upper plot of Fig.~\ref{fig:ex1_2IP_energies_num}. We found the expected non-renormalized degeneracies at $E=0,\pm4$ and three renormalized degeneracies labeled $d_0, d_{\pm 4}$. The extracted degeneracy values in each region are shown with black dots in Fig.~\ref{fig:ex1_2IP_energies_num}(a1-a3) for $U=1,2,3,4$. The comparison with the analytical (blue) curves reported in Fig.~\ref{fig:ex1_2IP_energies} shows excellent agreement. Moreover, as indicated by the numbers in the main plot of Fig.~\ref{fig:ex1_2IP_energies_num}, the non-renormalized degeneracies at $E=0,\pm4$ scale as $N^2$ while the renormalized energies scale as $N$, indicating that the renormalized compact states have macroscopic degeneracy.

\begin{figure}
    \centering
    \includegraphics[width=\columnwidth]{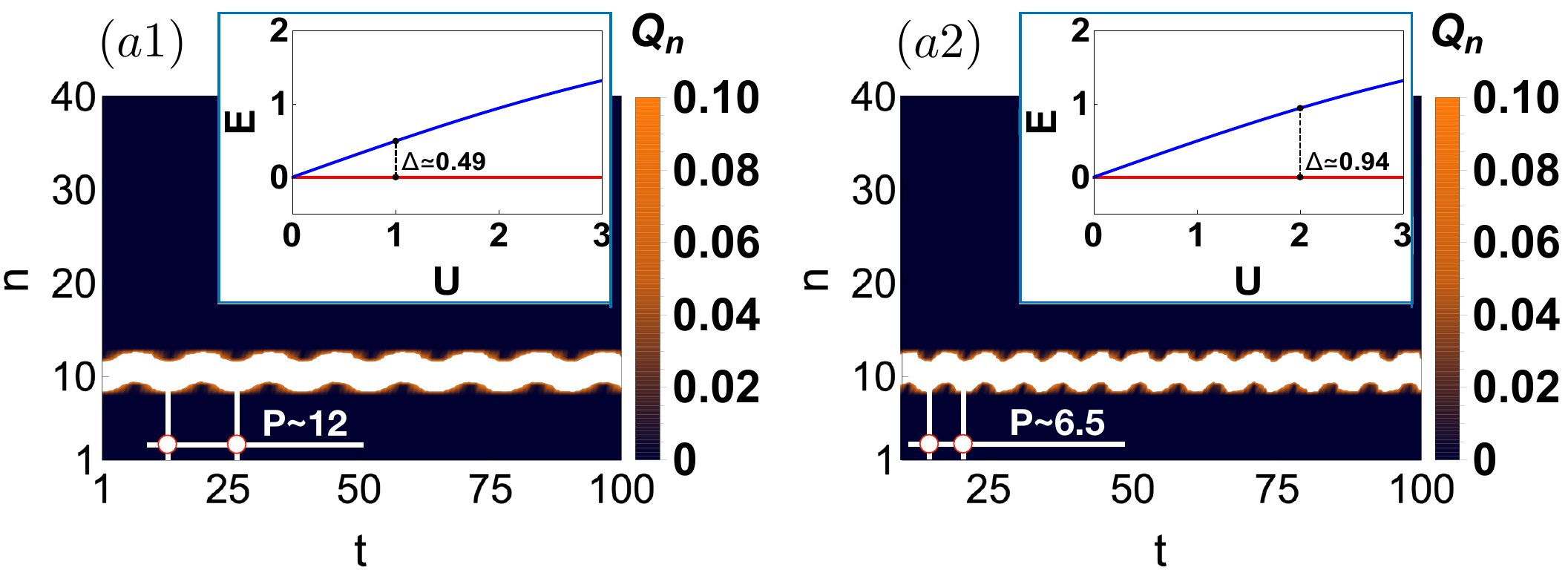}
    \caption{Time evolution of $Q_n$ for $U=1$ (a) and $U=2$ (b) with oscillation period $P$. Insets: zoom of Fig.\ref{fig:ex1_2IP_energies}(a1) and energy difference $\Delta$.
    The initial conditions are $E=0$ caged states of two non-interacting particles.}
    \label{fig:2IP_beating}
\end{figure}

These energy-renormalized states are beyond the grasp of the conserved quantities in Ref.~\onlinecite{tovmasyan2018preformed} and Sec.~\ref{sec:MBI} and they result in exact two particle quantum caging. Such exact caging manifests in {\it e.g.} beating of spatially compact excitations between non-renormalized and renormalized two particles states -- as shown in Fig.~\ref{fig:2IP_beating}.  
Numerically we compute the local density $\rho_{n,k}$ of the two particles and the corresponding one-dimensional PDF of the particle density $Q_n$ 
for $U=1$ (a1) and $U=2$ (a2). As initial condition we take $E=0$ caged states of two non-interacting particles. In both panels of Fig.~\ref{fig:2IP_beating}, we observe that the oscillation period $P$ (main plots; horizontal white measuring arrow) is related to the energy difference $\Delta$ (insets; vertical black measuring arrow) between renormalized and non-renormalized energies of caged states as $\Delta \approx \frac{2\pi}{P}$.

\subsubsection{Three particles and beyond}
\label{sec:M>=3}

The renormalized compact states of $M=2$ particles existing for Hamiltonian $\hmh$ in Eq.~(\ref{eq:BH1},\ref{eq:BH1_TBI}) under the condition  $|z_1|^2 =  |w_1|^2$ constitute the base case to show inductively the existence of such states for any finite number $3\leq M<\infty$ of interacting particles on an infinite lattice $N\rightarrow \infty$. The induction scheme unfolds as follows:
\begin{enumerate}
    \item The interaction matrix $V_M$ in Eq.~\eqref{eq:Mip_int_eqs} can be written as a sum of $M$ matrixes
    \begin{gather}
        V_M = \sum_{j=1}^M V_{M,n_{j}}.
    \label{eq:Mip_int_matrix}
    \end{gather}
    Each $V_{M,n_{j}}$ describes the interaction between $M-1$ particles, and it is obtained from $V_M$ by taking the $j^{th}$ particle as free (Appendix~\ref{app:int_VM_rec}). For any $1\leq j\leq M$, each interaction matrix $V_{M,n_{j}}$ acts on the $M-1$-dimensional networks obtained by fixing $n_j = const$ in the $M$-dimensional network Eq.~\eqref{eq:Mip_int_eqs};
    \item {\it Inductive hypothesis}: in each $M-1$-dimensional networks, the matrix $V_{M,n_{j}}$ yields renormalized compact states of  $M-1$ interacting particles. These states lie in components detangled from dispersive chains representing transporting states of paired particles - see Ref.~\onlinecite{tovmasyan2018preformed} and Sec.~\ref{sec:MBI};
    \item At the main diagonal of Eq.~\eqref{eq:Mip_int_eqs}, the detangled components encoding the renormalized compact states of $M-1$-particles merge, forming a larger detangled component. The $M-1$-particles compact states combine and yield compact states of $M$ particles whose energy is further renormalized.
\end{enumerate}

\begin{figure}
    \centering
    \includegraphics[width=\columnwidth]{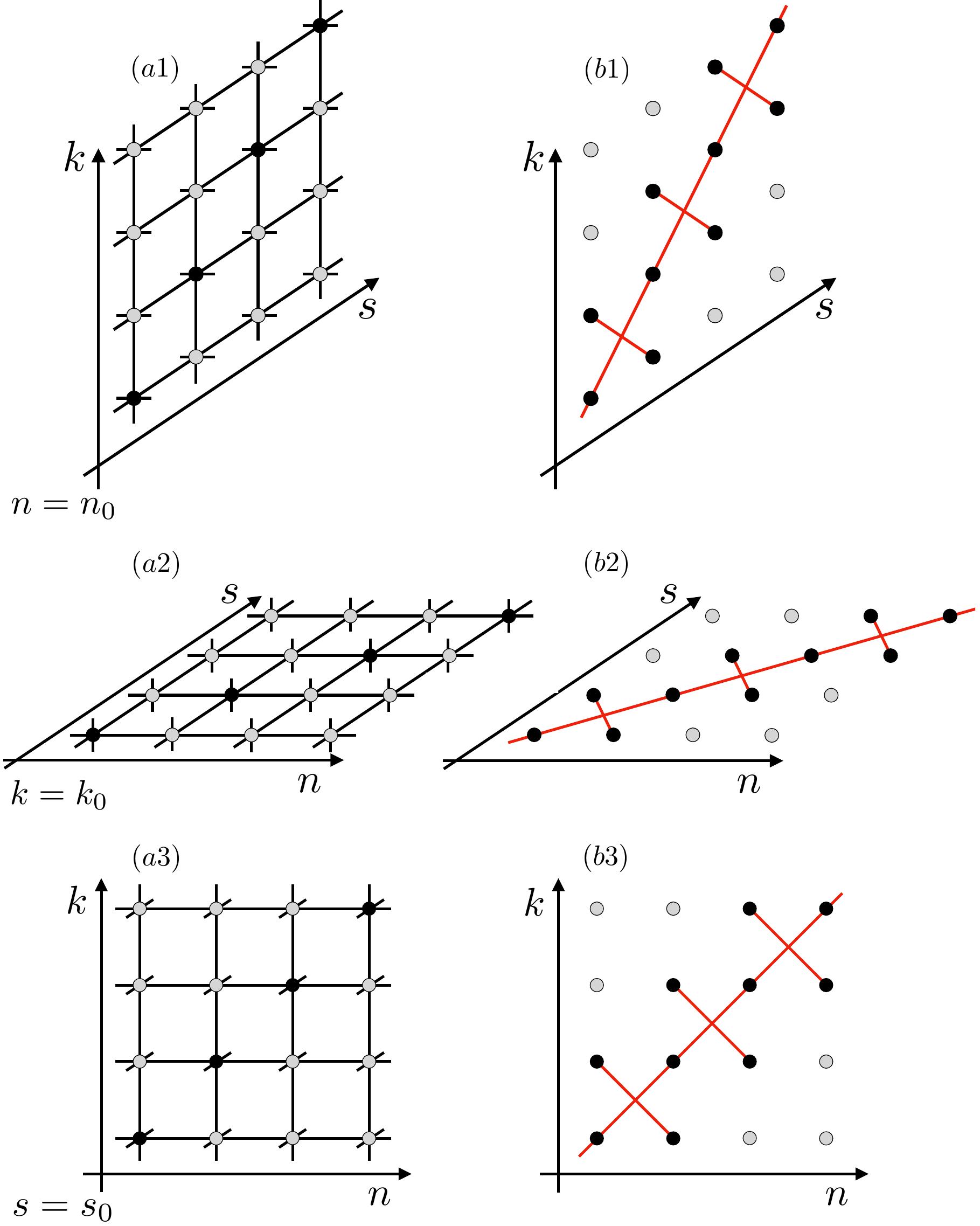}
    \caption{The graphical illustration of the decomposition given by Eq.~\eqref{eq:Mip_int_matrix_M=3}. 
    (a1) plane for $n=n_0$ of the 3D network Eq.~\eqref{eq:Mip_int_eqs} defined with $V_{3,n}$. Each dot indicates one unit cell $\varphi_{n,k,s}$, and the black lines are the hopping between cells. The black dots indicate the diagonal $k=s$ where $V_{3,n}$ applies (black dots). (b1) Rotated plane for $U\neq 0$ assuming $|z_1|^2 = |w_1|^2$. (a2-b2) and (a3-b3) are the same as (a1-b1) but for $k=k_0$ and $s=s_0$ respectively.}
    \label{fig:3p_2p}
\end{figure}

Let us visualize this inductive step for $M=3$ particles. In this case, Eq.~\eqref{eq:Mip_int_eqs} becomes a three dimensional network which can be fully detangled via a finite sequence of unitary transformations  for $U=0$. 
Following Eq.~\eqref{eq:Mip_int_matrix}, the interaction matrix $V_3$ in Eq.~\eqref{eq:Mip_int_eqs} can be decomposed as
\begin{align}
    V_3 &= V_{3,n} + V_{3,k} + V_{3,s}
    \label{eq:Mip_int_matrix_M=3}
\end{align}
Each matrix $V_{3,n}, V_{3,k} , V_{3,s}$ in Eq.~\eqref{eq:Mip_int_matrix_M=3} is reported in Appendix~\ref{app:int_VM_3} and is obtained from $V_3$ by considering one noninteracting particle and two interacting ones. Each matrix $V_{3,n}, V_{3,k} , V_{3,s}$ applies to a 2D network manifold (plane) of Eq.~\eqref{eq:Mip_int_eqs} obtained for $n=n_0$, $k=k_0$, and $s=s_0$ respectively -- as shown in Fig.~\ref{fig:3p_2p}(a1-a3). 

We consider each matrix in Eq.~\eqref{eq:Mip_int_matrix_M=3} separately, starting with $V_{3,n}$. In this case, along the diagonal $k=s$ of each plane  $n=n_0$, the matrix $V_{3,n}$ induces a one-dimensional chain Eq.~\eqref{eq:network_diag}. For the fine-tuned Hamiltonian $\hmh$ with $|z_1|^2 = |w_1|^2$ in Eq.~(\ref{eq:BH1},\ref{eq:BH1_TBI}), the one-dimensional chain Eq.~\eqref{eq:network_diag} decouples into a dispersive part with additional detangled  components -- as shown in Fig.~\ref{fig:3p_2p}(b1). Identical outcomes follow considering $V_{3,k}$ and $V_{3,s}$ -- as shown in Fig.~\ref{fig:3p_2p}(b2,b3). Indeed, along the diagonals $s=n$ ($s=k$) in each plane $k=k_0$ ($s=s_0$), the matrix $V_{3,k}$ ($V_{3,s}$) induces one-dimensional chains Eq.~\eqref{eq:network_diag} which decouple for the fine-tuned Hamiltonian $\hmh$. 

\begin{figure}
    \centering
    \includegraphics[width=0.95\columnwidth]{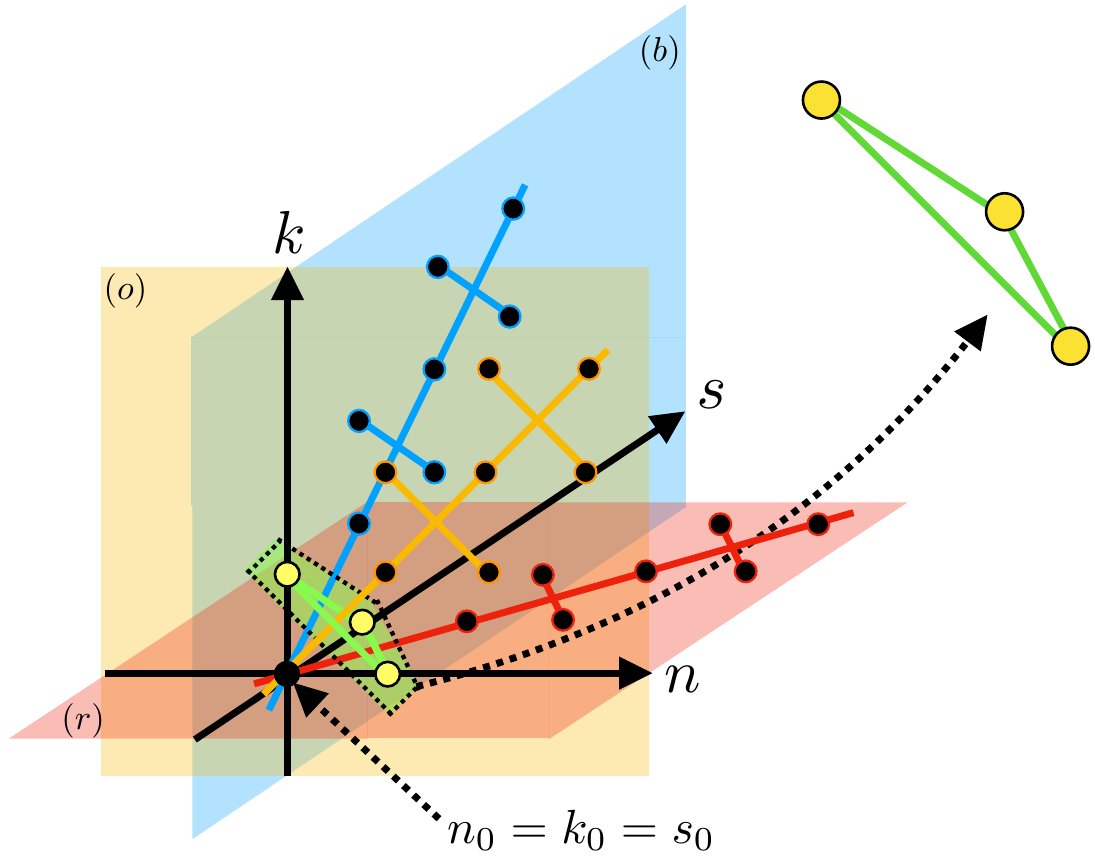}
    \caption{Diagonal structures of the rotated 3D network given by Eq.~\eqref{eq:Mip_int_eqs} for $M=3$ particles defined with the interaction matrix $V_{3}$. The detangled component with the compact states of three interacting particles is indicated with yellow dots and light green lines (also in the right top corner).}
    \label{fig:3p_CLS}
\end{figure}

Since Eq.~\eqref{eq:Mip_int_eqs} is linear, the resulting network induced by $V_3$ is obtained by combining the networks induced by  $V_{3,n}, V_{3,k} , V_{3,s}$ respectively -- shown in Fig.~\ref{fig:3p_2p}(b1-b3). The result is presented in Fig.~\ref{fig:3p_CLS}. In each plane $n=n_0$ [blue(b)], $k=k_0$ [red(r)], $s=s_0$ [orange(o)] the one-dimensional detangled chain Eq.~\eqref{eq:network_diag} along the respective diagonal is shown. Away from the main diagonal $n_0=k_0=s_0$  of Eq.~\eqref{eq:Mip_int_eqs}, the obtained dispersive chains yield extended transporting bound states corresponding to paired particles freely evolving along the chain, with the third unpaired particle remaining caged - as predicted in Ref.~\onlinecite{tovmasyan2018preformed}. The detangled components within each plane represent compact states of two interacting particles with the third particle caged away from the bound pair -  dubbed "\emph{2IP+1}" states. At the main diagonal $n_0=k_0=s_0$ of Eq.~\eqref{eq:Mip_int_eqs}, the detangled components of each plane form a unique component -- highlighted as yellow dots and light green lines and zoomed in the right top corner of Fig.~\ref{fig:3p_CLS}. This triangular component is detangled from the dispersive chains of each plane, and it encodes compact states of three interacting particles -- states dubbed "\emph{3IP states}".

\begin{figure}
    \centering
    \includegraphics[width=0.95\columnwidth]{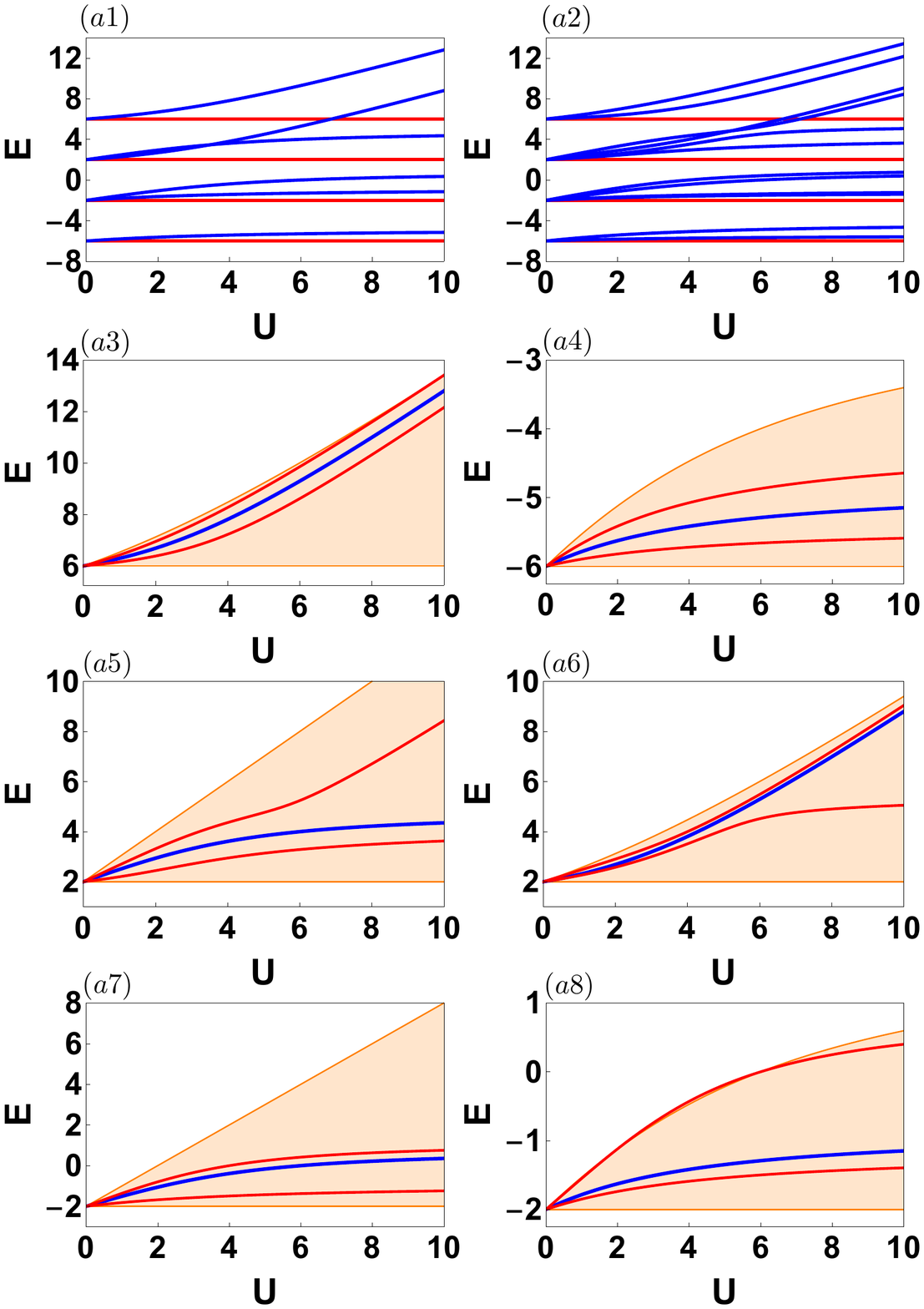}
    \caption{(a1) Non-renormalized degenerate energies (red) and renormalized energies (blue) of 2IP+1 states versus $U$. 
    (a2) Same as (a1) for 3IP states. 
    (a3-a8) Renormalized energies of 2IP+1 states (blue curve) and 3IP states (red curves) within the dispersive bands (orange areas) of 2IP states. }
    \label{fig:ex1_3IP_energies}
\end{figure}

To verify these general results for Hamiltonian system $\hmh$ fine-tuned as $|z_1|^2 = |w_1|^2$ in Eq.~(\ref{eq:BH1},\ref{eq:BH1_TBI}), we studied both the detangled components representing the  2IP+1 and the 3IP compact states for our test case. 
We found six renormalized energies of the 2IP+1 states as zeroes of $g_{F,2+1}(E;U)$ in the polynomial 
\begin{align}
    p_{F,2+1} & (E;U) = (E^2 - 36) (E^2 - 4)^4 \notag \\
    & \left[ E^3  - E^2 (U+6) + 4 E (U -1) +  4 (U+6)   \right] \notag \\
    & \left[ E^3  - E^2 (U-6) - 4 E (U +1) +  4 (U-6)   \right] \notag \\
    & \equiv (E^2 - 36) (E^2 - 4)^4  g_{F,2+1}(E;U)
    \label{eq:2+1p_int_offdiag_charpoly_app_V3s}
\end{align}
which also accounts for 10 non-renormalized degeneracies at $E=\pm2,\pm6$. We then found twelve renormalized energies of the 3IP states as zeroes of $ g_{F,3}(E;U)$ in the polynomial 
\begin{align}
    p_{F,3}&(E;U)= (E^2 -36) (E^2 -4)^5 \notag \\
    & \left[(E^2 -36 )^2 ( E^2 -4)^4  \right. \notag \\
    & \left.\  -    4 U E  (E^2 - 36) (E^2 -28) ( E^2 -4)^3  \right. \notag \\
    &  \left.\    +U^2 ( E^2 -4 )^2 (6 E^6 - 345 E^4 + 4888 E^2  -1872)  \right. \notag \\
    & \left. \ -  2 U^3 E (E^2 -4) (2 E^6  - 105 E^4 + 1376 E^2  -1392)  \right. \notag \\
    & \left.\  + U^4 (E^8  - 49 E^6 + 616 E^4  - 1168 E^2 + 576) \right] \notag \\
    & \equiv (E^2 -36) (E^2 -4)^5  g_{F,3}(E;U)
    \label{eq:3p_int_offdiag_charpoly_app_V3n_V3s}
\end{align}
which also accounts for 24 non-renormalized degeneracies at $E=\pm2,\pm6$. These renormalized energies are shown with blue curves in Fig.~\ref{fig:ex1_3IP_energies}(a1) and (a2) respectively. In the other six panels Fig.~\ref{fig:ex1_3IP_energies}(a2-a8), we plot each renormalized degeneracy of 2IP+1 states shown in panel (a1) in blue color, the renormalized degeneracy of 3IP states shown in panel (a2) in red color, and one of the six dispersive bands with orange shaded areas -- the latter ones obtained as zeroes of $g_{D,2+1}$ in the polynomial 
\begin{align}
    p_{D,2+1}&(E,k;U) = (-4 + E^2) \notag \\
    & \left[ ( E^2 -36 ) (E^2-4)^2 - 2 E (E^4 - 32 E^2 + 112) U\right. \notag \\
    & \left. + (E^4 - 24 E^2  + 48 ) U^2   + 32 U^2\cos k \right. \notag \\
    & \left. - 16 (E^2 -4  ) (E - U) U \cos k \right] \notag \\
    & \equiv (-4 + E^2)  g_{D,2+1}(E,k;U) 
    \label{eq:2+1p_int_disp_charpoly_app_V3s}
\end{align}
We observe that, as in the two particles case, all the renormalized energies lie within one dispersive bands, characterizing these renormalized compact states as quantum three particle bound states in the continuum (BIC).~\cite{hsu2016bound}~
\footnote{In Fig.\ref{fig:ex1_3IP_energies}(a3) and (a6) one of the two 3IP renormalized energies approaches the boundary of the continuum} 

\begin{figure}
    \centering
    \includegraphics[width=\columnwidth]{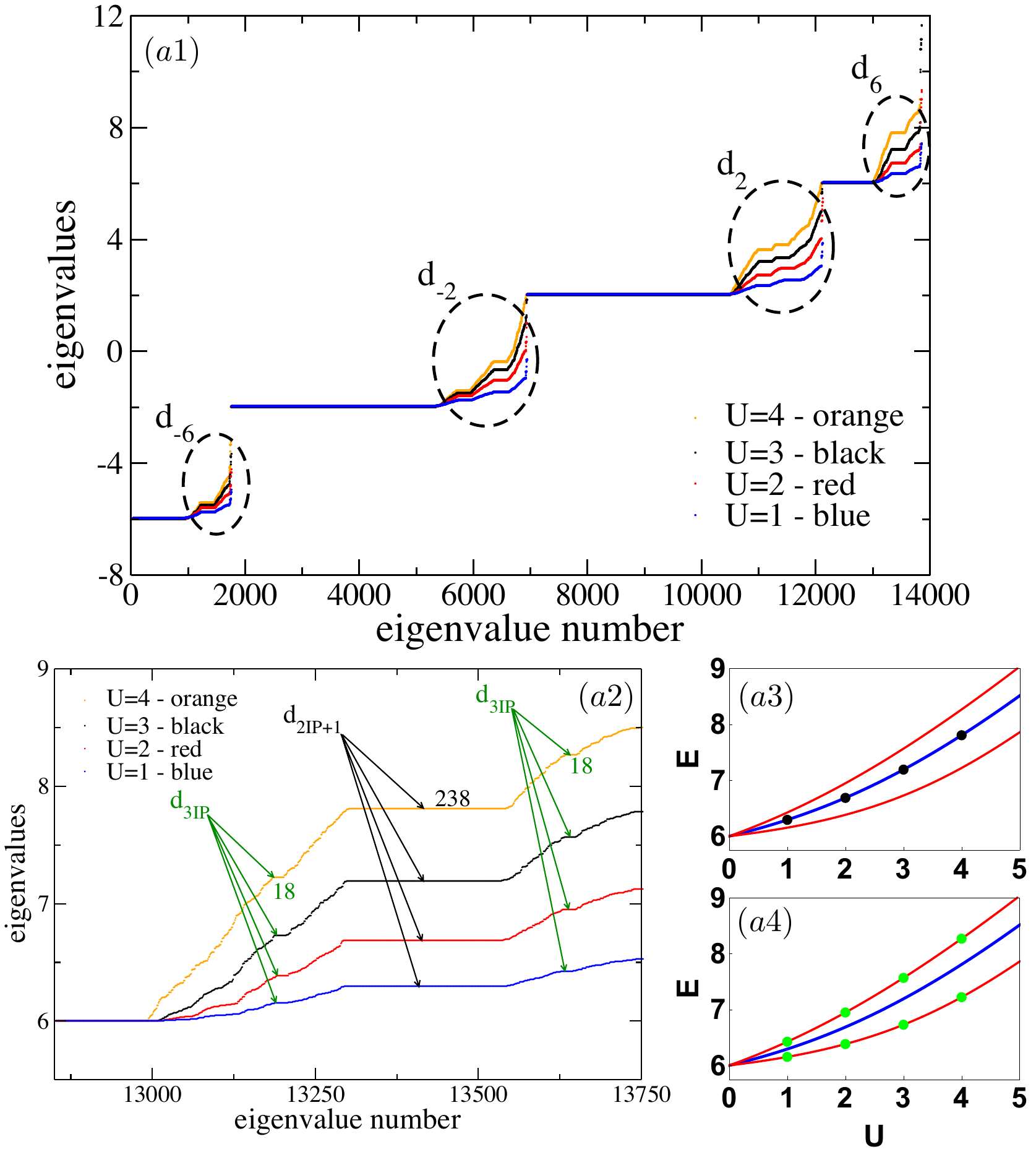}
    \caption{(a1) Eigenvalues of $M=3$ interacting bosons for $N=12$. (a2) Zoom on $d_6$. The numbers indicate the degeneracies of eigenvalues. (a3-a4) Renormalized energies of 2IP+1 states (blue curve) and 3IP states (red curves) versus $U$ and numerically obtained degeneracies - black dots (a3) and green dots (a4) respectively.}
    \label{fig:ex1_3IP_energies_num}
\end{figure}

Similarly to the two particles states, we numerically tested these results by diagonalizing the three dimensional network Eq.~\eqref{eq:Mip_int_eqs} associated to $M=3$ bosons for $U=1,2,3,4$ and $N=12$ unit cells. The results are reported in Fig.~\ref{fig:ex1_3IP_energies_num}. We found the expected non-renormalized degeneracies at $E=\pm2,\pm6$ and four renormalized degeneracies labeled $d_{\pm2}, d_{\pm 6}$. Let us focus on one of the four regions, {\it e.g.} $d_6$, zoomed in panel (a2). For each $U$, we observed three different degeneracies: one indicated with black arrows, and two indicated with green arrows. As shown in panel (a3), the former degeneracies  (black arrows) show excellent agreement with the analytical curve reported in Fig.~\ref{fig:ex1_2IP_energies}(a1) for 2IP+1 compact states. In this case, the degeneracy level is proportional to $2 N^2$, which follows from the fact that while 2IP have degeneracy N, the additional caged single particle can be placed in $\sim 2 N$ sites. As shown in panel (a4), the latter degeneracies (green arrows) show  excellent agreement with the analytical curve reported in Fig.~\ref{fig:ex1_3IP_energies}(a2) for 3IP compact states. In this case, the degeneracy level is proportional to $2 N$ sites, indicating that also the renormalized 3IP compact states have macroscopic degeneracy. 



As an inductive conjecture, this scheme can be repeated for any finite number $4\leq M < \infty $ of interacting particles by splitting the matrix $V_M$ split as sum of $M$ separate matrices as in Eq.~\eqref{eq:Mip_int_matrix}. For {\it e.g.} $M=4$ particles,  there exist: (i) renormalized compact states of three interacting particles plus one caged away (3IP+1 states); (ii) renormalized compact states made by two separate clusters of two interacting particles (2IP+2IP states); (iii) renormalized  compact states of four interacting particles (4IP states). This indicates that generically for $M$ particles there exist not only degenerate interaction renormalized compact states of $M$ interacting particles, but also renormalized compact states of $M$ particles formed as product states of renormalized compact states of $L\leq M$ particles.

\section{Conclusions and perspectives}

In this work we showed that the quantum version of classical nonlinear models exhibiting caging discussed in Ref.~\onlinecite{danieli2020cagingI} necessarily feature an extended set of conserved quantities -- number parity operators, firstly introduced in~\cite{tovmasyan2018preformed} for specific ABF geometries -- and transport is realized through moving pairs of interacting particles while single unpaired particles remain caged. 
We then demonstrated that the picture is more complex, as using $\nu=2$ ABF chain as testbeds we additionally showed the presence of energy renormalised multiparticle caged states -- states first observed for two fermions in AB diamond chain.~\cite{vidal2000interaction} We explicitly showed the existence of these macroscopically degenerate energy renormalized compact states for $M=2,3$ particles, and used an inductive conjecture to generalize them to any finite number $M < \infty $ of particles on an infinite lattice $N\rightarrow \infty$. Consequently, these caged many-body states generalize the extensively studied and experimentally observed single particles compact localized states~\cite{derzhko2015strongly,leykam2018artificial,leykam2018perspective} and hits towards quantum caging phenomena in interacting systems  - see {\it e.g.} Ref.~\onlinecite{diliberto2019nonlinear} for the AB diamond chain. 
Moreover, we showed that few particle $M=2,3$ caged states are bound states in the continuum (BIC)~\cite{hsu2016bound} -- as their renormalized energies reside within dispersive bands of delocalized states of pairs of particles. Conjecturing this for any finite number of particles $M$ yields that these systems constitute a potential platform for the experimental realization of BICs in quantum systems -- as discussed in Ref.~\onlinecite{hsu2016bound}.

The quantum caging of interacting particles highlights the all bands flat networks as a promising platform for novel phenomena in quantum many-body physics. We conjecture that these energy renormalised multiparticle caged states appear in any network whose classical version exhibits nonlinear caging. Confirmation of that hypothesis would lead to a systematic generalization of the quantum caging phenomenon in all band flat models to any number of bands $\nu\geq 3$ and higher spatial dimensions $d=2,3$. An important front is the fate of these states at the finite-density limit and their impact on the many-body dynamics. While the interaction in general allows paired particles to freely move along the system~\cite{tovmasyan2018preformed} forbidding MBL - a fact which can be overcome by fine-tuning the interaction~\cite{danieli2020many,kuno2020flat,orito2020exact} -- these states may contribute to anomalous thermalization phenomena. Thus the quest to search for anomalous many-body dynamics can not only be extended to interacting ABF lattices including disorder, dissipation and external fields, but also to lattice Hamiltonian supporting both flat and dispersive bands.~\cite{tilleke2020nearest}

\section{Acknowledgments}

The authors thank Ihor Vakulchyk, Ajith Ramachandran, Arindam Mallick and Tilen \v{C}adez for helpful discussions. This work was supported by the Institute for Basic Science, Korea (IBS-R024-D1).


\appendix

\section{Rotating the quantum interaction Hamiltonian $\hmhb_1$}
\label{app3b} 

The Hubbard interaction Hamiltonian $\hmhb_1$ in Eq.~\eqref{eq:BH1_TBI} for $\nu=2$ reads
\begin{align}
    \hmhb_1 &= \frac{U}{2} \sum_{n\in\mathbb{Z}} \left[ \ha_{n}^{\dagger}\ha_{n}^{\dagger}\ha_{n}\ha_{n} + \hb_{n}^{\dagger}   \hb_{n}^{\dagger}\hb_{n}\hb_{n} \right].
    \label{eq:BH1_TBI_app}
\end{align}
The unitary transformation that transforms the two band ladder Eqs.~(\ref{eq:H0H1_rot0},\ref{eq:H0H1_rot1}) into decoupled dimers 
written for the operators 
$\ha_{n}, \hb_{n}$ and $\hp_{n}, \hf_{n} $ reads 
\begin{gather}
    \begin{cases}
        \ha_n = e^{i\theta_1}\left(z_1 \hp_n + w_1 \hf_n \right) \\
        \hb_n = e^{i\theta_1} \left(-w_1^* \hp_n +  z_1^*\hf_n  \right) \\   
    \end{cases}
    \label{eq:ex1_trU1_c_app}
\end{gather}
Then, $ \hmhb_1$ in Eq.~\eqref{eq:BH1_TBI_app} turns to
\begin{align}
    \hmhb_1 &= \frac{U}{2}\sum_n   \left\{  ( |z_1|^4 +  |w_1|^4) \hp_n^\dagger\hp_n^\dagger \hp_n \hp_n  \right.  \notag \\
    &\left.\qquad +2 z_1^*  w_1 (|z_1|^2 - |w_1|^2)  \hp_n^\dagger\hp_n^\dagger \hp_n \hf_n  \right. \notag \\
    &\left. \qquad+ 2 z_1^{*2} w_1^{2} \hp_n^\dagger\hp_n^\dagger \hf_n \hf_n  \right.  \notag \\
    \label{eq:BH_int_a+b_app}
    &\left.\qquad+ 2 z_1  w_1^* ( |z_1|^2  - |w_1|^2)  \hp_n^\dagger \hf_n^\dagger \hp_n \hp_n  \right. \\
    &\left.\qquad+8 |z_1|^2 |w_1|^2  \hp_n^\dagger \hf_n^\dagger \hp_n \hf_n  \right.  \notag \\
    &\left. \qquad+ 2 z_1^* w_1 (|w_1|^2 - |z_1|^2) \hp_n^\dagger \hf_n^\dagger \hf_n \hf_n  \right.  \notag \\
    &\left.\qquad+ 2 z_1^{2} w_1^{*2} \hf_n^\dagger \hf_n^\dagger \hp_n \hp_n  \right. \notag \\
    &\left.\qquad+ 2 z_1 w_1^* (|w_1|^2 -  |z_1|^2 )  \hf_n^\dagger \hf_n^\dagger \hp_n \hf_n  \right. \notag \\
    &\left. \qquad + ( |z_1|^4 + |w_1|^4)  \hf_n^\dagger \hf_n^\dagger  \hf_n \hf_n \right\} \notag
\end{align}
The fine-tuning $|w_1|^2 = |z_1|^2$ simplifies Eq.~\eqref{eq:BH_int_a+b_app} to 
\begin{align}
    \hmhb_1 & = U\sum_n  \left\{  |z_1|^4   \left[ \hp_{n}^{\dagger} \hp_{n}^{\dagger} \hp_{n} \hp_{n} + \hf_{n}^{\dagger} \hf_{n}^{\dagger} \hf_{n} \hf_{n} + 4 \hp_{n}^{\dagger} \hf_{n+1}^{\dagger} \hp_{n} \hf_{n+1} \right] \right. \notag \\
    &\left. \quad  + z_1^{*2} w_1^2 \hp_{n}^\dagger \hp_{n}^\dagger \hf_{n+1}\hf_{n+1} +  z_1^{*} w_1^{*2} \hp_{n} \hp_{n} \hf_{n+1}^\dagger\hf_{n+1}^\dagger \right\} 
    \label{eq:Ham1_BH_app}
\end{align}

\section{$M$ interacting particles}
\label{app:Mdim_sch}

\subsection{Mapping an $M$-particles problem to an $M$-dimensional  Schr\"odinger system}
\label{app:mapping_M}

Let us consider the wave function $\ket{\psi}$ of $M$ interacting particles evolving on the ABF network described by the Hamiltonian  $\hmhb=\hmhb_0+\hmhb_1$ in Eqs.~(\ref{eq:BH1},\ref{eq:BH1_TBI}). Then, we consider the multi-index ${\mathbf n}=(n_1,\dots,n_M)$, and the entry $n_s$ indicates the unit cell where where the $s^{th}$ particles is located -- either on the $a$ or the $b$ chain. 
Hence, for a given multi-index ${\mathbf n}\in\mathbb{Z}^M$ there exist a $2^M$-dimensional vector of elements of the Fock basis $\ket{v_{\mathbf n}}_{\bf a,b} = \{\ket{v_{\mathbf n}^s}_{\bf a,b} \}_{s\leq 2^M}$, representing all possible particles configurations. 
%
%

The wave function $\ket{\psi}$ is expanded as
\begin{gather}
    \ket{\psi} = \sum_{{\mathbf n}\in\mathbb{Z}^M} \varphi_{\mathbf n} \cdot \ket{v_{\mathbf n}}_{\bf a,b}
    \label{eq:MIP_wf}
\end{gather}
for $\varphi_{\mathbf n}$ a $2^M$-dimensional complex vector.\\
\\ 
\noindent
Then, proceed as follows:
\begin{enumerate}
    \item substitute Eq.~\eqref{eq:MIP_wf} in the equation $i \partial_t \ket{\psi} =  \hmhb \ket{\psi}$ with $\hmhb=\hmhb_0+\hmhb_1$ in Eqs.~(\ref{eq:BH1},\ref{eq:BH1_TBI});
    \item unfold the products in the r.h.s. of  $i \partial_t \ket{\psi} =  \hmhb \ket{\psi}$. Then regroup those terms with common index $n_s$ and common element of the basis  $\ket{v_{\mathbf n}^s}_{\bf a,b}$
    \item multiply the obtained equation by ${_{\bf a,b}}\langle v_{\mathbf n}^s |$. This yields the equation correspondent to the $s^{th}$ component of the vector $\varphi_{\mathbf n}$
    \item group these equations in vector frm for $\varphi_{\mathbf n}$. 
\end{enumerate}
    
\noindent    
This results in the $M$-dimensional Schr\"odinger system 
\begin{gather}
    i \dot{\varphi}_{\mathbf n} = \left[A +  U V_M \right]  \varphi_{\mathbf n} + \sum_{j=1}^M \left[ T_j \varphi_{{\mathbf n} + {\mathbf e}_j}  + T_j^\dagger \varphi_{{\mathbf n} - {\mathbf e}_j}\right] 
    \label{eq:Mip_int_eqs_app}
\end{gather}
where $ {\mathbf e}_j$ are the canonical basis of $\mathbb{Z}^M$. The matrices $A,\{T_j\}_{j\leq M}$ describe the lattice geometry induced by $\hmhb_0$, while $V_M$ is a diagonal matrix encapsulating the interaction $\hmhb_1$ 
whose entrees $v_s$ correspond to the $s^{th}$ element $\ket{v_{\mathbf n}^s}_{\bf a,b}$ of the chosen Fock basis in Eq.~\eqref{eq:MIP_wf} 
\begin{gather}
    v_s = {_{\bf a,b}}\bra{v_{\mathbf n}^s} \hmhb_1 \ket{v_{\mathbf n}^s}_{\bf a,b}
    \label{eq:VM_vj}
\end{gather}
For $U=0$ and any number of particles $M$, Eq.~\eqref{eq:Mip_int_eqs_app} has $2^M$ flatbands.

\subsection{Two Interacting Particles Case}
\label{app_2IP}

For $M=2$, Eq.~\eqref{eq:Mip_int_eqs_app} reduces to a two dimensional system of Schr\"odinger equations
\begin{align}
    i \dot{\varphi}_{n,k} =  \left[A +  U V_2 \right]  \varphi_{n,k} 
    & + T_n \varphi_{n+1,k} + T_n^\dagger \varphi_{n-1,k} \notag \\
    & + T_k \varphi_{n,k+1} + T_k^\dagger \varphi_{n,k-1}
    \label{eq:2p_int_eqs_app}
\end{align}
with $\varphi_{n,k} = (X_{n,k},Y_{n,k},W_{n,k},Z_{n,k})^T$. The onsite matrixes $A$ is 
\begin{align}
    A  & = 2 \Gamma_0
    \begin{pmatrix}
        |z_1|^2- |w_1|^2 & -z_1w_1 & -z_1w_1 & 0 \\
        -z_1^*w_1^* & 0  & 0 & -z_1w_1 \\
        -z_1^*w_1^* & 0 & 0 & -z_1w_1 \\ 
        0 & -z_1^*w_1^* & -z_1^*w_1^* & |w_1|^2-|z_1|^2  \\ 
    \end{pmatrix}   
    \label{eq:2p_int_matrx1_app}
\end{align}
while the hopping matrixes $T_n,T_k$ are
\begin{align}
    \label{eq:2p_int_matrx2n_app}
    T_n  &=  \Gamma_1
    \begin{pmatrix}
        z_1  w_1^* & 0 &  z_1^2  & 0 \\
        0 & z_1  w_1^*  & 0 &  z_1^2  \\
        - (w_1^*)^2 & 0 & -z_1  w_1^* & 0 \\ 
        0 & - (w_1^*)^2 & 0 & -z_1  w_1^* \\ 
    \end{pmatrix}     \\                      
         %
    T_k   &=  \Gamma_1
    \begin{pmatrix}
        z_1  w_1^* & z_1^2 & 0 & 0 \\
        -(w_1^*)^2 & -z_1  w_1^*  & 0 & 0 \\
        0 & 0 & z_1  w_1^* & z_1^2 \\ 
        0 & 0 & -(w_1^*)^2 & -z_1  w_1^* \\ 
    \end{pmatrix}
    \label{eq:2p_int_matrx2k_app}
\end{align}
for $\Gamma_0 =  |w_2|^2 -  |z_2|^2$ and $\Gamma_1 = 2 z_2 w_2$. 
The matrix $V_2$ encoding the interaction reads
\begin{align}
    V_2  & = 
    \begin{pmatrix}
        \delta_{n,k} & 0 & 0 & 0 \\
        0 & 0  & 0 & 0 \\
        0 & 0 & 0 & 0 \\ 
        0 & 0 & 0 & \delta_{n,k} \\ 
    \end{pmatrix}.
    \label{eq:2p_int_matrx3_app}
\end{align}
For $U=0$, Eq.~\eqref{eq:2p_int_eqs_app} has four flatbands: one at energy $E= 2$ and one $E= -2$; and two at $E=0$.

\subsection{Three Interacting Particles Case}
\label{app_3IP}

For $M=3$, Eq.~\eqref{eq:Mip_int_eqs_app} reduces to a three dimensional system of Schr\"odinger equations
\begin{align}
    i \dot{\varphi}_{n,k,s} =   \left[A +  U V_3 \right]  \varphi_{n,k,s} &+ T_n \varphi_{n+1,k,s}  + T_n^\dagger \varphi_{n-1,k,s} \notag \\
    &+ T_k \varphi_{n,k+1,s}  +T_k^\dagger \varphi_{n,k-1,s} \notag \\
    &+ T_s \varphi_{n,k,s+1}  + T_s^\dagger \varphi_{n,k,s-1}.
    \label{eq:3p_int_eqs_app}
\end{align}
for the eight-component vector 
\begin{align}
    \varphi_{n,k,s} = ( & X_{n,k,s}, Y^{(1)}_{n,k,s} ,  Y^{(2)}_{n,k,s},  W^{(3)}_{n,k,s}, \notag \\
    & W^{(1)}_{n,k,s} ,  W^{(2)}_{n,k,s},  W^{(3)}_{n,k,s} , Z_{n,k,s})^T.
    \label{eq:3p_vect}
\end{align}
We skip the cumbersome matrixes $A,T_n,T_k,T_s$, and only report the matrix $V_3$ encoding the interaction
\begin{align}
    V_3  & =
     \footnotesize
    \begin{pmatrix}
        \Delta_{n,k,s} & 0 & 0 & 0 & 0  & 0 & 0 & 0 \\
        0 & \delta_{n,k}  & 0  & 0  & 0 & 0  & 0 & 0 \\
        0  & 0 & \delta_{n,s}  & 0  & 0  & 0 & 0  & 0\\
        0 & 0  & 0 & \delta_{k,s} & 0 & 0 & 0 & 0 \\ 
        0  & 0 & 0 & 0 & \delta_{k,s} & 0  & 0  & 0 \\ 
        0 & 0  & 0  & 0  & 0 & \delta_{n,s}  & 0 & 0\\
        0  & 0 & 0  & 0  & 0  & 0 &\delta_{n,k}  & 0 \\
        0  & 0  & 0  & 0 & 0  & 0  & 0 & \Delta_{n,k,s}
    \end{pmatrix}   
    \label{eq:3p_int_matrx_V3}
\end{align}
where for $\Delta_{n,k,s} = \delta_{n,k} + \delta_{n,s} + \delta_{k,s}$.
For $U=0$, Eq.~\eqref{eq:3p_int_eqs_app} has eight flatbands: one at energy $E= 3$ and one $E= -3$; three at $E=1$ and three at $E=-1$.

\subsection{Detangling the $M$-dimensional Schr\"odinger system for noninteracting particles}
\label{app:detangling_M}

The single particle Hamiltonian $\hmhb_0$ in Eq.~\eqref{eq:BH1} can be fully detangled via a sequence of 3 unitary transformations (two rotations and one unit-cell redefinition - Ref.~\onlinecite{danieli2020cagingI}). As a consequence, this holds also for its associated Schr\"odinger system Eq.~\eqref{eq:Mip_int_eqs_app} for $U=0$. In this case the detangling procedure consists in a sequence of $3 M$ unitary transformations, where each triplet of transformations consists in two unitary rotations and one unit cell redefinition. Each triplet of coordinate redefinitions recursively zeroes one of the hopping matrix $T_j$ and redefine the intra-cell matrix $A$. After $M$ triplets of unitary transformations all hopping matrixes $\{T_j\}_{j\leq M}$ vanish, and the matrix $A$ turn diagonal with the $2^M$ flatband energies as diagonal entrees.

By applying this procedure, the two dimensional system Eq.~\eqref{eq:2p_int_eqs_app} is equivalent to a fully detangled network
\begin{align}
    i \dot{\varphi}_{n,k}  
     &=  A_6 \varphi_{n,k} \ , 
     \qquad
         A_6  =  
    \begin{pmatrix}
        -2 & 0 & 0 & 0 \\
        0 & 0  & 0 & 0 \\
        0 & 0 & 0 & 0 \\ 
        0 & 0 & 0 & 2 \\ 
    \end{pmatrix}. 
    \label{eq:2p_int_eqs_rot_U1}
\end{align}
where the diagonal elements of $A_6$ in Eq.~\eqref{eq:2p_int_eqs_rot_U1} are the flatband energies.

Likewise, the three dimensional system Eq.~\eqref{eq:3p_int_eqs_app} is ultimately equivalent to a fully detangled network
\begin{align}
    i \dot{\varphi}_{n,k,s}
    & =  A_9 \varphi_{n,k,s}.
    \label{eq:3p_int_eqs_rot_U1}
\end{align}
The diagonal elements of the onsite matrix $A_9$ in Eq.~\eqref{eq:3p_int_eqs_rot_U1} are the flatband energies. 

\section{Renormalized compact states of two interacting particles}
\label{app2}

\subsection{Lattice of Fano defects}
\label{app5_fano}

The one-dimensional chain shown in Fig.~\ref{fig:ex2_2IP}(a2) along the three main diagonals $k=n,n\pm 1$ of the rotated Eq.~\eqref{eq:2p_int_eqs_app}
\begin{gather}
    i \dot{\eta}_n = K_D \eta_n +  H_D  \eta_{n+1} +  H_D^\dagger  \eta_{n+1}
    \label{eq:network_maindiag}
\end{gather} 
for the unit cell $\eta_n = (\varphi_{n-1,n} , \varphi_{n,n} , \varphi_{n,n-1})$ 
is defined by the block matrixes $K_D,H_D$ which depend on the interaction strength $U$
\begin{gather}
    K_D = 
    \begin{pmatrix}
        V^{(n-1,n)} & U \bar{T}_{n}^{(n-1,n)} & U B \\
        U\bar{T}_{n6}^{(n-1,n)\dagger} & V^{(n,n)}   & U \bar{T}_{k}^{(n,n-1)} \\
        U B^\dagger   & U \bar{T}_{k}^{(n,n-1)\dagger} & V^{(n,n-1)} 
    \end{pmatrix},
    \label{eq:matrx_maindiag_MD}
\end{gather}
\begin{gather}
    H_D = U
    \begin{pmatrix}
        \mathbb{O}_4 & \mathbb{O}_4  & \mathbb{O}_4  \\
        \mathbb{O}_4   & C  & \bar{T}_{n}^{(n,n+1)} \\
        \mathbb{O}_4   &  \mathbb{O}_4  & \mathbb{O}_4 
    \end{pmatrix}
    \label{eq:matrx_maindiag_HD}
\end{gather}

\noindent
We omit the cumbersome definitions of the blocks defining $K_D,H_D$.

It turns out that the hopping matrix $\bar{T}_{k}^{(n,n-1)}, \bar{T}_{n}^{(n-1,n)},  \bar{T}_{n}^{(n,n+1)}$ have $\xi = z_1 w_1^* (|z_1|^2 - |w_1|^2)$ as a common prefactor.
The fine-tuning condition $|z_1|^2 = |w_1|^2$ yields $\xi=0$ and the matrixes in Eqs.~(\ref{eq:matrx_maindiag_MD},\ref{eq:matrx_maindiag_HD}) reduce to
\begin{gather}
    K_D = 
    \begin{pmatrix}
        V^{(n-1,n)} &  \mathbb{O}_4 & U B \\
        \mathbb{O}_4 & V^{(n,n)}   &  \mathbb{O}_4 \\
        U B^\dagger   &  \mathbb{O}_4 & V^{(n,n-1)} 
    \end{pmatrix}
    \label{eq:matrx_maindiag_MD_xi0}
\end{gather} 
\begin{gather}
    H_D = U
    \begin{pmatrix}
        \mathbb{O}_4 & \mathbb{O}_4  & \mathbb{O}_4  \\
        \mathbb{O}_4   & C  &  \mathbb{O}_4 \\
        \mathbb{O}_4   &  \mathbb{O}_4  & \mathbb{O}_4 
    \end{pmatrix}
    \label{eq:matrx_maindiag_HD_xi0}
\end{gather}
and the network in Eq.~\eqref{eq:network_maindiag} reduces to a lattice of Fano-defects~\cite{flach2014detangling}: it decouples in a $4$-components dispersive chain
\begin{gather}
    i \dot{\varphi}_{n,n} =  V^{(n,n)} \varphi_{n,n} +U\left[ C \varphi_{n+1,n+1}  + C^\dagger \varphi_{n-1,n-1}\right]
    \label{eq:matrx_maindiag_1Dchain}
\end{gather} 
and a detangled (Fano) part
\begin{gather}
    i \frac{\partial}{\partial t}
    \begin{pmatrix}
        \varphi_{n-1,n} \\
        \varphi_{n,n-1} \\  
    \end{pmatrix}  
    = 
    \begin{pmatrix}
        V^{(n-1,n)}   & U B \\
        U B^\dagger  & V^{(n,n-1)} 
    \end{pmatrix}
    \begin{pmatrix}
        \varphi_{n-1,n} \\
        \varphi_{n,n-1} \\  
    \end{pmatrix}  
    \label{eq:matrx_maindiag_Fano}
\end{gather}
shown in Fig.~\ref{fig:2D_mapped}(b4) of the main text.

\subsection{An example}
\label{app5_CLS_ex1}

In the case of the test case obtained with $z_i = \cos \pi/4 $, $w_i = \sin \pi/4$ for $i=1,2$ in Eqs.~(\ref{eq:H0H1_rot0},\ref{eq:H0H1_rot1}), the onsite matrixes $V_7^{(n,n)} ,V_7^{(n-1,n)}, V^{(n,n-1)} $ read
\begin{align}
    V^{(n,n)} &= \frac{1}{4}
    \begin{pmatrix}
        U-16 & 0 & 0 & U \\
        0 & U  & U & 0 \\
        0 & U & U & 0 \\ 
        U & 0 & 0 & U+16 \\ 
    \end{pmatrix},  
    \label{eq:ex1_V7nn}
\end{align}
\begin{align}
    V^{(n-1,n)} &=  \frac{1}{8}
    \begin{pmatrix}
        U-32 & -U & U & -U \\
        -U & U  & -U & U \\
        U & -U & U & -U \\ 
        -U & U & -U & U+32 \\ 
    \end{pmatrix},       
    \label{eq:ex1_V7nn1}
\end{align}
\begin{align}
    V^{(n,n-1)} &=  \frac{1}{8}
    \begin{pmatrix}
        U-32 & U & -U & -U \\
        U & U  & -U & -U \\
        -U & -U & U & U \\ 
        -U & -U & U & U+32 \\ 
    \end{pmatrix},                                    
    \label{eq:ex1_V7n1n}
\end{align}
while the matrixes $B,C$ read
\begin{gather}
    B =  \frac{1}{8}
    \begin{pmatrix}
        1 & 1 & -1 &- 1 \\ 
        - 1 & -1 & 1 & 1 \\ 
        1 & 1 & -1 & -1 \\
        -1 & -1  & 1 & 1 
    \end{pmatrix},
    \quad       
    C = \frac{1}{8}
    \begin{pmatrix}
        1 & 1 & 1 & 1 \\ 
        -1 & -1 & -1 & -1 \\ 
        -1 & -1 & -1 & -1 \\
        1 & 1  & 1 & 1 
    \end{pmatrix}.      
    \label{eq:2p_cross_matrix_Un0}
\end{gather}

\noindent
The dispersive component Eq.~\eqref{eq:matrx_maindiag_1Dchain} has a characteristic polynomial
\begin{align}
    p_D(E,k;U) & = E  [E^3 - U E^2  -16 E + 8 U(1-\cos k) ] \notag\\
    & \equiv E \, g_D(E,k;U) 
    \label{eq:2p_int_disp_charpoly_app}
\end{align}
The detangled component Eq.~\eqref{eq:matrx_maindiag_Fano} has characteristic polynomial  
\begin{align}
    p_F(E;U) & = E^3 ( E^2 -16 ) (E^3 - U E^2  -16 E + 8 U ) \notag \\
    & \equiv E^3 ( E^2 -16 ) \cdot g_F(E;U)
    \label{eq:2p_int_offdiag_charpoly_app}
\end{align}

\section{Interaction matrix $V_M$ decomposition}
\label{app:int_VM}

\subsection{Recursive formula}
\label{app:int_VM_rec}

If from $M$ interacting particles we form $M$ groups of $M-1$ interacting particles by taking the $j^{th}$ as free, then from a state $\ket{v_{\mathbf n}^s}_{a,b}$ we get $M$ states $\ket{v_{\mathbf n}^{s,j}}_{a,b}$. 
It then follows the set of $M$ coefficients  
\begin{gather}
    v_s^j = {_{a,b}}\bra{v_{\mathbf n}^{s,j}}\hmhb_1\ket{v_{\mathbf n}^{s,j}}_{a,b}
    \label{eq:VM_vj_s}
\end{gather}
Let us suppose that all $M\geq 3$ particles are on chain $a$ (likewise, chain $b$). It then follows 
\begin{gather}
    v_s = \sum_{t=1}^M \sum_{r=t }^M \delta_{n_t,n_r}\ ,
    \qquad
    v_s^j = \sum_{\substack{t=1 \\ t\ne j}}^M \sum_{r=t}^M \delta_{n_t,n_r}
    \label{eq:int_matrix_id1}
\end{gather}
since the Kroneker delta involving $n_j$ are excluded (the $j^{th}$ particle is free).
This yields that the sum of $v_s^j$ is
\begin{gather}
    \sum_{j=1}^M v_s^j =  (M-2) v_s.
    \label{eq:int_matrix_id3}
\end{gather}
since any Kroneker delta $\delta_{n_t,n_r}$ in $ v_s$ is absent in $v_s^t$ and $v_s^r$ but it exists once in all the other $M-2$ terms $v_s^j$. 
The same relation in Eq.(\ref{eq:int_matrix_id3}) also hold for $L$ particles on chain $a$ and $M-L$ particles on chain $b$. 

We define the matrixes $V_{M,n_{j}}$ as diagonal matrixes whose diagonal elements are the coefficients $v_s^j$ in Eq.~\eqref{eq:VM_vj_s} divided by $M-2$. 
This yields the decomposition Eq.~\eqref{eq:Mip_int_matrix} for any number of particles $M\geq 3$
\begin{gather}
    V_M = \sum_{j=1}^M V_{M,n_{j}}
    \label{eq:Mip_int_matrix_app}
\end{gather}

\subsection{Three particles case}
\label{app:int_VM_3}

From the interaction matrix $V_3$ in Eq.~\eqref{eq:3p_int_matrx_V3}, we obtain the matrix
\begin{align}
    V_{3,n}  & =   
     \footnotesize
    \begin{pmatrix}
        \delta_{k,s} & 0 & 0 & 0 & 0  & 0 & 0 & 0 \\
        0 & 0  & 0  & 0  & 0 & 0  & 0 & 0 \\
        0  & 0 & 0  & 0  & 0  & 0 & 0  & 0\\
        0 & 0  & 0 & \delta_{k,s} & 0 & 0 & 0 & 0 \\ 
        0  & 0 & 0 & 0 & \delta_{k,s} & 0  & 0  & 0 \\ 
        0 & 0  & 0  & 0  & 0 & 0 & 0 & 0\\
        0  & 0 & 0  & 0  & 0  & 0 & 0  & 0 \\
        0  & 0  & 0  & 0 & 0  & 0  & 0 & \delta_{k,s}
    \end{pmatrix}.
    \label{eq:3p_ks_V3n}
\end{align}
by considering the particle indexed via $n$ non-interacting with the remaining two. Likewise, the matrix
\begin{align}
    V_{3,k}  & =   
         \footnotesize
    \begin{pmatrix}
        \delta_{n,s} & 0 & 0 & 0 & 0  & 0 & 0 & 0 \\
        0 & 0  & 0  & 0  & 0 & 0  & 0 & 0 \\
        0  & 0 & \delta_{n,s}  & 0  & 0  & 0 & 0  & 0\\
        0 & 0  & 0 &0  & 0 & 0 & 0 & 0 \\ 
        0  & 0 & 0 & 0 & 0 & 0  & 0  & 0 \\ 
        0 & 0  & 0  & 0  & 0 & \delta_{n,s}  & 0 & 0\\
        0  & 0 & 0  & 0  & 0  & 0 &0  & 0 \\
        0  & 0  & 0  & 0 & 0  & 0  & 0 & \delta_{n,s}
    \end{pmatrix}.
    \label{eq:3p_ns_V3k}
\end{align}
is obtained by considering the particle labeled by $k$ non-interacting with the other two, 
while the matrix 
\begin{align} 
    V_{3,s}  & =   
     \footnotesize
    \begin{pmatrix}
        \delta_{n,k} & 0 & 0 & 0 & 0  & 0 & 0 & 0 \\
        0 & \delta_{n,k}  & 0  & 0  & 0 & 0  & 0 & 0 \\
        0  & 0 & 0  & 0  & 0  & 0 & 0  & 0\\
        0 & 0  & 0 & 0 & 0 & 0 & 0 & 0 \\ 
        0  & 0 & 0 & 0 & 0 & 0  & 0  & 0 \\ 
        0 & 0  & 0  & 0  & 0 & 0 & 0 & 0\\
        0  & 0 & 0  & 0  & 0  & 0 &\delta_{n,k}  & 0 \\
        0  & 0  & 0  & 0 & 0  & 0  & 0 & \delta_{n,k}
    \end{pmatrix}. 
    \label{eq:3p_nk_V3s}
\end{align}
is obtained by considering the particle labeled by $s$ non-interacting with the other two. 
The sum of these three matrixes
\begin{align}
    V_3 &= V_{3,n}+V_{3,k} + V_{3,s}.
    \label{eq:Mip_int_matrix_M=3_app}
\end{align}
yields $V_3$ in Eq.~\eqref{eq:3p_int_matrx_V3} -- confirming the formula Eq.~\eqref{eq:Mip_int_matrix_app}.

\bibliography{flatband,mbl}

\end{document}